
%
%
%
%
%

\magnification=1200
\parskip=5pt plus 2pt
\vsize=9.1 true in
\hsize=6.2 true in
\parindent=20pt
\baselineskip=17pt


\newcount\notenumber
\def\resetnotenumber{\notenumber=1}
\def\note#1{{\baselineskip=12pt\footnote{$^{\the\notenumber}$}{#1}}
\advance\notenumber by1}
\resetnotenumber

\def\ie{{\it i.e.,\ }}

\def\nte#1{\baselineskip=12pt\footnote{$^{\ast}$}{#1}}


\def\R{{\rm I\! R}}
\def\\C{{\rm C}}
\def\C{\mkern1mu\raise2.2pt\hbox{$\scriptscriptstyle|$}{\mkern-7mu\rm C}}

\def\and{{\rm\ and\ }}

\def\abs #1{\vert {#1}\vert}
\def\frac#1#2{{#1 \over #2}}
\def\half{{\frac 12}}
\def\twid #1{\tilde {#1}}

\def\what #1{\widehat {#1}}

\def\Mir{{\int_{\Sigma} d^3 x \,}}

\def\Tr {\rm Tr}
\def\sproduct{{\ooalign{\hfil\raise.07ex\hbox{s}\hfil\crcr\mathhexbox20D}}}


\def\a {\alpha}
\def\b {\beta}
\def\g {\gamma}
\def\G {\Gamma}
\def\om {\omega}
\def\Om {\Omega}
\def\d {\delta}
\def\e {\epsilon}

\def\l {\lambda}

\def\vf {\chi}

%
%

\def\am {\cal A}

\def\em {\cal E}

\def\de {{\rm {det}}}

\def\Me {\cal M}
\def\On {1\over \Om}
\def\Si {\Sigma}
\def\te {\theta}
\def\Te {\Theta}
\def\bta {{\beta} _0}
\def\btb {{\beta} _+}
\def\btc {{\beta} _-}
\def\cet {\hbar}
\def\plu {\R ^+}
\def\Lig {{\cal L} ^+_{(+,+)}}
\def\w {\wedge}
\def\z {\phantom z}
%
%
{\nopagenumbers
\rightline {SU-GP-93/4-4, gr-qc/9304041}
\rightline {April 1993}
\vglue 0.5 in
\centerline {\bf Non-Perturbative Canonical Quantization of
Minisuperspace Models:}
\centerline {\bf Bianchi Types I and II}
\bigskip
\centerline {Nenad Manojlovi\'c}
\smallskip
\centerline {and}
\smallskip
\centerline {Guillermo A. Mena Marug\'an
\nte{On leave from Instituto de Matem\'aticas y F\' {\i}sica
Fundamental, C.S.I.C., Serrano 121, 28006 Madrid, Spain.}
}
\centerline {\it Physics Department, Syracuse University,}
\centerline {\it Syracuse, NY 13244-1130, USA.}
\medskip
\centerline {\bf Abstract}
\noindent
We carry out the quantization of the full type I and II Bianchi models
following the non-perturbative canonical quantization program. These
homogeneous minisuperspaces are completely soluble, \ie it
is possible to obtain the general solution to their classical equations
of motion in an explicit form. We determine the sectors of solutions
that correspond to different spacetime geometries, and prove that the
parameters employed to describe the different physical solutions
define a good set of coordinates in the phase space of these models.
Performing a transformation from the Ashtekar variables to this set
of phase space coordinates, we endow the reduced phase space of
each of these systems with a symplectic structure. The symplectic
forms obtained for the type I and II Bianchi models are then identified
as those of the cotangent bundles over ${\cal L} ^+_{(+,+)}
\times S ^2 \times S ^1$ (modulo some identification of points)
and ${\cal L} ^+_{(+,+)} \times S ^1$,
respectively, with ${\cal L} ^+_{(+,+)}$ the positive quadrant of
the future light-cone. We construct a closed \break
${\ast} -$ algebra
of Dirac observables in each of these reduced phase spaces,
and complete the quantization program by
finding unitary irreducible representations of these algebras.
The real Dirac observables are represented in this way by self-adjoint
operators, and the spaces of quantum physical states are provided
with a Hilbert structure.

\eject}

\pageno=2
\noindent
{\bf I. Introduction}
\medskip
\noindent
Quantization of gravity is one of the main obstacles that Modern
Physics has to face in order to obtain an unified treatment of all physical
interactions. To achieve this goal, a new and promising approach, the
non-perturbative canonical quantization program, has been developed
systematically over the last years [1-3]. Even though this approach has
succeeded in solving a variety of physical problems [3],
the implementation of this quantization program for the full theory of
gravity remains incomplete.

	To apply the canonical quantization program to a given theory,
one first selects an over-complete set of complex classical variables
in the phase space of the system that is closed under
the Poisson-brackets structure [4]. This set is promoted to an abstract
${\ast} -$ algebra of quantum elementary operators, with the complex
conjugation relations between classical variables translated into
${\ast} -$ relations in this algebra. The next step consists of
choosing a complex vector space, and finding on it a representation of
the ${\ast}-$ algebra of elementary operators.
The kernel of all the operators that represent the constraints
of the system provides us with the space of quantum solutions to all
physical constraints in the considered representation, \ie with
the space of quantum states. At this stage of the
quantization program, one should determine a set of ``real''
Dirac observables for the theory, that is, a set of
operators which correspond to real classical variables and such that they
commute weakly with all the quantum constraints [4]. If this set is
sufficiently large, one can fix uniquely the inner product in
the quantum physical space by requiring that the ``real'' Dirac observables
are promoted to self-adjoint operators (reality conditions) [2,3,5].
The space of quantum states is endowed in this way with a Hilbert
structure. Finally, to extract predictions from the
quantum theory so constructed, one has to supply the obtained
mathematical framework with a physical interpretation.

	Owing to the great complexity of the general theory of gravitation,
both the complete space of physical states and a satisfactory set of
Dirac observables for gravity are still to be determined in the new
variables formalism. The analysis and quantization of
minisuperspace gravitational models, following the non-perturbative
canonical program, can be at this point a helpful way to develop
some insight into the kind of problems, methods, and techniques
that are involved in the quantization of the full theory of gravity.
In addition to the lessons one can learn by applying
the quantization program to simple models, it is clear that
obtaining consistent quantum theories for minisuperspace models
of cosmological interest is physically relevant by itself, as it enables
us to address cosmological problems quantum mechanically. It is therefore not
surprising that the recent literature contains
a considerable number of works on
canonical quantization of minisuperspace models [6-11].

	A special attention has been paid, in particular, to the quantization
of the Bianchi models [7-10]. These are spatially homogeneous spacetimes
which admit an isometry group that acts transitively on each leaf of the
homogeneous foliation [12,13]. Nevertheless, the analysis on
Bianchi models, has been restricted almost entirely to the case
of diagonal models [9-11], \ie models in which the metric is purely diagonal.
Even if this reduction is completely consistent [11], it would be desirable to
carry out the whole quantization program for the full non-diagonal Bianchi
models. In this way, one could also study the role played by the extra
non-diagonal degrees of freedom in the quantum version of these systems,
and discuss the implications of the diagonal reduction from the quantum point
of view.

	In this paper, we will analyze in detail the full non-diagonal
type I and II Bianchi models. These systems are
completely soluble, that is, one can obtain the explicit expressions
of the general classical solution for both of these models. We will
prove that the parameters that appear in the general solution define
a good coordinatization of the phase space in these two models. The
use of the non-diagonal degrees of freedom turns out to be decisive in
determining the ranges of the introduced phase space coordinates. In
fact, the reduction to the corresponding diagonal models would lead us
to different conclusions about the range of the coordinates that
describe the diagonal degrees of freedom. This is essentially due
to the fact that, in the diagonal case, one can consistently consider the
spatial directions in the homogeneous foliation as fixed once and
forever. In the non-diagonal case, however, the requirement of
analyticity in the introduced coordinatization of the phase space
obliges us to deal exclusively with different homogeneous
three-geometries, which are invariant under any interchange
of the spatial directions that are not preferred by the symmetries
of the model.

	On the other hand, using the explicit expressions of
the classical solutions, it is possible to endow the
reduced phase space of each of these models with an
analytic symplectic structure. The symplectic forms
obtained in this way can be
interpreted as those associated to real cotangent bundles over some
specific reduced configuration spaces. Following the canonical
quantization program, we construct, on each of these reduced phase
spaces, an over-complete set of classical variables that commute weakly
with the constraints of the system and form a closed Lie algebra
with respect to the Poisson-brackets structure. This Lie algebra
can be identified as the algebra of the Dirac observables
of the theory. We will then choose a vector space and find on it an
unitary irreducible representation of the algebra of observables,
so that the real Dirac observables are represented by self-adjoint
operators [14]. The Hilbert spaces determined by this procedure provide
us with the spaces of quantum states of the studied models.

	The outline of this paper is as follows. In Sec. II we introduce
the class of mini\-superspace models on which we will concentrate in this
work, and the main formulas needed to carry out our analysis in
the new variables
formalism. The general classical solutions for
type I and II Bianchi models are obtained in Sec.III.
In Sec. IV we study the symplectic structure of the space of physical
solutions for Bianchi type II. The symplectic form in this space is
written in terms of the different parameters contained
in the general solution. We have to prove then that
the chosen set of parameters defines a good coordinatization of
the phase space of the model. This is the subject of Sec. V.
Our analysis for Bianchi II is generalized to the non-diagonal type I
Bianchi model in Sec. VI. Sec. VII deals with the non-perturbative
canonical quantization of these two models.
In Sec. VIII we present a different approach to the
quantization of Bianchi type I, using the symmetries that are
present in this model at the classical level to find a complete
set of generalized ``plane waves'' which span the space of
quantum states. We also show that the two quantum theories
constructed for type I are unitarily equivalent.
Finally, we summarize the results in
Sec. IX, where we also include some further discussions.

\bigskip

\noindent
{\bf II. Bianchi Models}
\medskip
\noindent
Bianchi models are spatially homogeneous spacetimes (\ie they can be
foliated by three dimensional Riemannian manifolds) which admit
a three dimensional isometry Lie group G that acts simply transitively
on each leaf $\Si$ of the homogeneous foliation [12,13]. As a consequence,
there exists for each of these models a set of three left-invariant
vector fields $L _I$ on $\Si$ which form the Lie algebra of the
group $G$:
$$
\eqalignno {
[ L _I , L _J ] &= {C ^K} _{IJ} L _K 		\, \, \, , 	&(2.1) \cr}
$$
where ${C ^I} _{JK}$ are the structure constants of the Lie group.
Dual to the the vector fields $L _I$, one can introduce a set of
three left-invariant one-forms $\vf ^I$ which satisfy the
Maurer-Cartan equations
$$
\eqalignno {
d {\vf} ^I + \half {C^I} _{JK} \, \, \, \vf ^J \w \vf ^K &= 0
\, \, \, .							&(2.2) \cr}
$$

	If the trace ${C ^I} _{IJ}$ of the structure constants is equal to
zero, the Bianchi model is said to belong to Bianchi class {\bf A}.
For this class of models, the spacetime admits foliations by
compact slices [15]. We will restrict ourselves to this case hereafter.

	The structure constants for the class {\bf A} Bianchi models
can always be written in the form [12,16]
$$
\eqalignno {
{C ^I} _{JK} &= \e _{JKL} S ^{LI}	\, \, \, ,		&(2.3) \cr}
$$
with $\e _{JKL}$ the anti-symmetric symbol, and $S ^{IL}$ a symmetric
tensor. Further classification of the class {\bf A} Bianchi models
is defined with respect to the signature of the symmetric tensor $S^{IJ}$
[12,16]. The type I and II Bianchi models, the only ones we will consider in
this work, are characterized by the signatures $( 0 , 0 , 0 )$ and
$( 0 , 0 , + )$, respectively. Thus, the structure constants for
Bianchi I are given by ${C ^I} _{JK} = 0$, while for Bianchi II
${C ^I} _{JK} = \d ^I_3 \e _{3JK}$.

	In the new variables formalism, one starts by introducing the
triads ${e ^i} _a (x)$ on a three-manifold $\Si$, where $i=1,2,3$
is a spatial index and $a= (1),(2),(3)$ is an $SO(3)$ vector index
which is raised and lowered  with the metric $\eta _{ab} = ( 1 , 1 , 1 )$.
The inverse metric on $\Si$ can be written in terms of the triads as
$g^{ij} = {e ^i} _a \, e ^{ja}$. The Ashtekar variables
$( {{\em} ^i} _a , {{\am} _j} ^b )$ are defined as
the densitized triad and the spin connection [3]:
$$
\eqalignno {
{{\em} ^i} _a = ({\de} g) ^{\half} {e ^i} _a \, \, \, ,& \quad
{{\am} _i} ^a = {\G _i} ^a (e) - {\rm i} \, {K _i} ^a
						\, \, \, ,&(2.4) \cr}
$$
with ${\G _i} ^a$ the $SO(3)$ connection compatible with ${e^i} _a$, and
${K _i} ^a$ the triadic form of the extrinsic curvature.

	For the class {\bf A} Bianchi models and $\Si$
a compact manifold, one can always perform
the following transformation of variables [17]:
$$
\eqalignno {
\bigl ( {{\am} _i} ^a ( x , t ) , {{\em} ^i} _a ( x , t ) \bigr ) & \to
\bigl ( {A _I} ^a (t) , {\twid {\am} _i} ^a ( x , t ) , {E ^I} _a (t) ,
{\twid {\em} ^i} _a ( x , t ) \bigr )				&(2.5) \cr}
$$
where $t$ is the time introduce by the homogeneous foliation, $x$ is a set
of coordinates on the leaf $\Si$, and
$$
\eqalignno {
{A _I} ^a  &= {\On} \Mir \abs {\vf} \, {{\am} _i} ^a {L _I} ^i
							\, , \quad
{{\twid {\am}} _i} ^a = {{\am} _i} ^a - {A _I} ^a {\vf ^I} _i
							 \, , 	&(2.6) \cr
{E ^I} _a &= {\On} \Mir {{\em} ^i} _a {\vf ^I} _i
							 \, , \quad
{{\twid {\em}} ^i} _a = {{\em} ^i} _a - {E ^I} _a {L _I} ^i | {\vf} |
							  \, .	&(2.7) \cr}
$$
In Eqs. (2.6,7), ${\vf} ^I = {\vf ^I} _i dx ^i$,
$L _I = {L _I} ^ i \partial _i$,
$| {\vf} |$ is the determinant of ${\vf ^I} _i$ and \break
$\Om = \int _{\Si} d ^3 x   | {\vf} |$. The reduction of
the dynamical degrees of freedom in the class {\bf A} Bianchi models
is accomplished by imposing
${{\twid {\am}} _i} ^a = {{\twid {\em}} ^i} _a = 0$.
In this way, one is left only with a finite number of degrees of freedom,
given by $( {A _I} ^a (t)  , {E ^I} _a (t) )$. The variables
$( {A _I} ^a (t)  , {E ^I} _a (t) )$ form a canonical set [11,17]:
$$
\eqalignno { \{ {A _I} ^a (t)  , {E ^J} _b (t) \} &= {\rm i} \d ^J_I \d ^a_b
						\, \, \, . 	& (2.8) \cr}
$$
In terms of them, the constraints for the class {\bf A} Bianchi models
adopt the following expressions [17]:
$$
\eqalignno {
{\cal G} _a &= {\e _{ab}} ^c {A _I} ^b {E ^I}_c \, \, \, ,	&(2.9) \cr
{\cal V} _I &= {C ^K} _{IJ} {A _K} ^a {E ^J} _a \, \, \, ,	&(2.10) \cr
{\cal S} &= {\e _a} ^{bc} \bigl ( - {C ^K} _{IJ} {A _K} ^a +
{\e ^a} _{de} {A _I} ^d {A _J} ^e \bigr ) \, {E ^I} _b {E ^J} _c
						\, \, \, ,	&(2.11) \cr}
$$
where ${\cal G} _a$, ${\cal V} _I$, and ${\cal S}$ denote, respectively,
the Gauss law, the vector constraints and the scalar constraint.

	On the other hand, using the left-invariant one-forms ${\vf} ^I$,
the spatial three-metric on $\Si$ can be written as
$g = g _{IJ} {\vf} ^I {\vf} ^J$, with ${E ^I} _a$ related to the inverse of
$g _{IJ}$ by means of
$$
\eqalignno { g ^{IJ} ( \de E ) &= {E ^I} _a {E ^J} ^a
						\, \, \, , 	&(2.12) \cr}
$$
and $\de E$ denoting the determinant of ${E ^I} _a$.

	Finally, the connection ${A _I} ^a$ admits the decomposition
$$
\eqalignno { {A _I} ^a &= {\G _I} ^a (E) - {\rm i} \, {K _I} ^a
                        			\, \, \, , 	&(2.13) \cr}
$$
where the $SO (3)$ connection ${\G _I} ^a$ and the triadic extrinsic curvature
${K _I} ^a$ can be determined through the formulas
$$
\eqalignno { {\G _I} ^a &= {{\e ^a} _b} ^c \bigl ( {\half} {C ^M} _{JI}
{(E ^{-1}) ^b} _M {E ^J} _c - {1\over 4} {C ^M} _{JI} {(E ^{-1}) ^d} _M
{(E ^{-1}) ^d} _I {E ^J} _b {E ^L} _c \bigr )
						\, \, \, , 	&(2.14) \cr
{K _I} ^a &= K _{IJ} {E ^J} _a (\de E) ^{- \half}
                                                \, \, \, , 	&(2.15) \cr}
$$
with ${(E ^{-1}) ^a} _I$ the inverse of ${E ^I} _a$ and $K _{IJ}$ the
extrinsic curvature associated to the metric $g _{IJ}$.

\noindent{\bf III. Bianchi Types I and II: Classical Solutions}
\medskip
\noindent
In the rest of this work we will restrict our attention to the type
I and II Bianchi models [12,18]. It is well known that, for these models,
one can always reduce the geometrodynamic initial value problem to the
diagonal case [19,20]. We will now briefly review the argument that leads
to this conclusion.

	Suppose that, for either of these two models, we begin by considering
a certain set of left-invariant one-forms ${\vf} ^I$, for which the
three-metric is given by $g_{IJ}$. Any other set of left-invariant forms
${\twid {\vf}} ^I$ will be related to ${\vf} ^I$ by a transformation
${\twid {\vf}} ^I = {M ^I} _J {\vf} ^J$ that maintains the symmetries
of the model. The structure constants must, therefore,
remain unchanged under the transformation defined by ${M ^I} _J$,
$$
\eqalignno { {C ^I} _{JK} &= {( M ^{-1} ) ^I} _L {C ^L} _{PQ}
{M ^P} _J {M ^Q} _K 				\, \, \, . 	&(3.1) \cr}
$$
The inverse $( M ^{-1} )$ must always exist, since ${\twid {\vf}} ^I$
is a set of three linearly independent one-forms.

	For Bianchi type I, ${C ^I} _{JK} = 0$, and condition (3.1) is empty.
In this case, any invertible matrix $M \in GL (3,\R)$ defines a permissible
transformation. Thus, for any geometrodynamic initial value data
$( {g ^0} _{IJ} , {K ^0} _{IJ} )$ in the set ${\vf} ^I$ (with ${g ^0} _{IJ}$
a positive definite metric), we can perform a transformation with a matrix
$M \in {GL (3,\R)\over SO(3)}$ such that, in the new set of one-forms,
the initial metric takes the value ${{\twid g} ^0} _{IJ} = \d _{IJ}$.
Then, using a transformation
under $SO(3)$, we can bring the initial extrinsic curvature ${K ^0} _{IJ}$
to the diagonal form, without altering the identity value for the initial
metric [19,20]. Since the diagonal ansatz is compatible with the dynamics of
the type I Bianchi model, we conclude as a corollary that any
geometrodynamic classical solution for type I
can be expressed as
$$
\eqalignno { g _{IJ} &= {(M ^t) _I} ^D g _D {M ^D} _J
						\, \, \, ,	&(3.2) \cr}
$$
where the metric $g _D$ is a classical solution for the diagonal case,
${M ^I} _J$ is a constant invertible matrix, and $(M ^t)$ denotes
the transpose of $M$.

	Let us consider now the type II Bianchi model. For this model,
${C ^I} _{JK} = \d ^I_3 \e _{3JK}$, and condition (3.1) implies
that ${M ^I} _J$ must be of the form
$$
\eqalignno { M &=
    \left ( \matrix { {\cal M} & \matrix { 0 \cr 0 \cr} \cr
\matrix { M ^3_1 & M ^3_2 \cr} &           {\de } {\Me} \cr} \right )
					\, \, \, , 	&(3.3) \cr}
$$
with ${\Me} \in GL (2,\R)$. Given any initial value data
$( {g ^0} _{IJ} , {K ^0} _{IJ} )$, defined in the set of one-forms
${\vf} ^I$, we can always carry out a transformation under a matrix
of the type (3.3), with ${\Me} \in {GL (2, \R)\over SO(2)}$, such that,
in the new set ${\twid {\vf}} ^I = {M ^I} _J {\vf} ^J$, the initial metric
is equal to ${{\twid g} ^0} _{IJ} = \d _{IJ} + ({{\twid g} ^0} _3 - 1 )
\d _{I3} \d _{J3}$, and the extrinsic curvature is \break
${{\twid K} ^0} _{IJ} = {(M ^t) _J} ^L {{\twid K} ^0} _{LN} {M ^N} _J$.

	Using expressions (2.12-15), the Gauss law constraints
(2.9) and the two non-empty vector constraints in (2.10) for Bianchi
type II, it is possible to see that, for positive definite metrics
${{\twid g} ^0} _{IJ}$ of the form that we have obtained,
${{\twid K} ^0} _{31}$ and ${{\twid K} ^0} _{32}$ must vanish if
$( {{\twid g} ^0} _{IJ} , {{\twid K} ^0} _{IJ} )$ is an admissible set
of initial value data. With these conditions, it is clear that
${{\twid K} ^0} _{IJ}$ can be brought to
diagonal form by a transformation of the type (3.3) with ${\Me} \in S0(2)$
and ${M ^3} _1= {M ^3} _2 = 0$ [19]. Under such a transformation,
the initial metric
${{\twid g} ^0} _{IJ}$ remains unchanged. Since the diagonal case is consistent
with the dynamics of the model, we conclude, as for Bianchi I, that any
Bianchi type II classical solution can be written in the form (3.2),
with $M$ an invertible constant matrix of the type (3.3).

	Therefore, to get the general solution in geometrodynamics for
the Bianchi types I\break
and II, it suffices to find the classical
solutions for the corresponding diagonal cases. These solutions can
in fact be obtained from the analysis of the diagonal Bianchi models
made by Ashtekar, Tate and Uggla in Ref. [9]. Parallelling their
notation, we introduce the following parametrization for the
diagonal metric $g _D$, $D = 1, 2, 3$:
$$
\eqalignno { g _1 = e ^{2{\sqrt 3} ( {\bta} - {\btb} + {\btc} )} \, ,
\quad g _2 &= e ^{2{\sqrt 3} ( {\bta} - {\btb} - {\btc} )} \, ,
\quad g _3 = e ^{2{\sqrt 3} {\btb}} \, .			&(3.4) \cr}
$$
In the Misner's gauge [21], defined by the lapse function
$N = 12 \, (\de g_D ) ^{\half}$ \break
$= 12 \, e ^{{\sqrt 3} (2{\bta} - {\btb})}$,
the dynamical equations for Bianchi type I adopt the simple expression
$$
\eqalignno { {\dot \bta} = - p _0	\, , \quad  {\dot \btb} &= p _+ \, ,
\quad {\dot \btc} = p _- 				\, ,	&(3.5) \cr}
$$
where the dot denotes time derivative and $(p _0, p _+, p _-)$ are
three real constants which are related through the scalar constraint [9]
$$
\eqalignno { {\cal S} & \propto - {p_0} ^2 + {p_+} ^2 + {p_-} ^2 = 0
						\, \, \, .	&(3.6) \cr}
$$
Integrating the equations of motion, we arrive at a diagonal metric
and a lapse \break
function
\note{The integration constants for $\b _0 , \b _+$ and $\b _-$
can be absorbed by a translation of the origin of time and a
redefinition of the matrix M that appears in (3.2).}
$$
\eqalignno { g _1 = e ^{-2{\sqrt3} ( p_0 + p_+ - p_- ) t}
							\, , \quad
g _2 &= e ^{-2{\sqrt3} ( p_0 + p_+ + p_- ) t}
							\, , \quad
g _3 = e ^{2{\sqrt3} p_+ t}
							\, ,	&(3.7) \cr
N &= 12 e ^{ - {\sqrt3} ( 2 p_0 + p_+ ) t}
						\, \, \, . 	&(3.8) \cr}
$$
Finally, from (3.6), we can restrict our analysis to non-negative
$p_0$ given by
$$
\eqalignno { p_0 &= \sqrt{ {p_+} ^2 + {p_-} ^2} \, \, \, .	&(3.9) \cr}
$$
The solutions corresponding to negative $p_0$ can be obtained from those
with $p_0 > 0$ by changing the sign of the time parameter $t$ that
defines the evolution and flipping the signs of $p_+$ and $p_-$. All the
different physical solutions (\ie solutions with different spacetime
geometries) are contained in the sector
$p_0 \geq 0, p_+,p_-,t \in \R$ [9].
We will thus restrict ourselves to this range of the parameters appearing in
(3.7-9).

	To find the classical solution for Bianchi type II, one needs to
perform a canonical transformation that mixes $g _3$ with its canonical
momentum [9]:
$$
\eqalignno { e ^{2{\sqrt 3} {\btb}} =
{{\bar p _+}\over {2{\sqrt 3} \cosh (2{\sqrt3} {\bar \btb})}} \, ,  &\quad
p _+ = - {\bar p_+} \tanh ( 2{\sqrt 3} {\bar \btb} )	\, , 	&(3.10) \cr}
$$
with ${\bar p_+}$ defined as a strictly positive variable. In Misner's
gauge, the dynamical equations (3.5) and the scalar constraint (3.6) are
still valid for the type II diagonal model, with the substitution $({\btb} ,
p_+) \to ( {\bar \btb} , {\bar p _+} )$, and $(p _0, {\bar p _+}, p _-)$
three real constants. It is then straightforward to derive the general
expression for the diagonal metric in the classical solutions
$$
\eqalignno { g _1 &= e ^{-2{\sqrt 3} ( p_0 - p_- ) t} \,
{{2{\sqrt3} \cosh (2{\sqrt 3} {\bar p_+} t )}\over {\bar p_+}}
						\, \, \, ,	&(3.11a) \cr
             g _2 &= e ^{-2{\sqrt 3} ( p_0 + p_- ) t} \,
{{2{\sqrt3} \cosh (2{\sqrt 3} {\bar p_+} t )}\over {\bar p_+}}
						\, \, \, ,	&(3.11b) \cr
	     g _3 &= {{\bar p_+}\over {2{\sqrt 3}
\cosh (2{\sqrt 3} {\bar p_+} t )}}		\, \, \, ,	&(3.11c) \cr}
$$
where $t$ is the time coordinate defined by the lapse function
$$
\eqalignno { N &= 12 e ^{-2{\sqrt 3} p_0 t} \,
\Bigl ( {{2{\sqrt3} \cosh (2{\sqrt 3} {\bar p_+} t )}\over {\bar p_+}}
\Bigr ) ^{\half} 				\, \, \, , 	&(3.12) \cr}
$$
and
$$
\eqalignno { {\bar p _+} > 0 \, , \quad p_- \in \R \, , &\quad
p_0 = \sqrt{ {p_+} ^2 + {p_-} ^2} > 0 			\, .	&(3.13) \cr}
$$

	Using the explicit expressions (3.7,8) and (3.11,12) for the metric
and the lapse function in the diagonal type I and II Bianchi models, the
relations (2.12-15) and (3.2), and the formula
$$
\eqalignno { K _{IJ} &= {1\over 2N} {\dot g} _{IJ} \, \, \, , 	&(3.14) \cr}
$$
valid in the gauge in which the shift functions vanish [13],
one can easily compute
the general form of the Bianchi I and II triads and spin connections in the
physical solutions, restricted to the sector that corresponds to positive
definite metrics. The result can be written in the compact notation
$$
\eqalignno { {E ^I} _a &= \de M {(M ^{-1}) ^I} _D E _D {\cal R} ^{aD}
						\, \, \, , 	&(3.15) \cr
             {A _I} ^a &= {(M ^t) _I} ^D \, {{\om} _D\over {E _D}} \,
{\cal R} ^{aD}					\, \, \, ,	&(3.16) \cr}
$$
where ${\cal R} ^{aD}$ is a general complex orthogonal matrix, and the sum over
$D = 1, 2, 3$ is implicitly assumed.

	For Bianchi type I, the matrix $M$ belongs to $GL(3,\R)$ and
$$
\eqalignno { E _1 &= e ^{-{\sqrt3} ( p_0 + p_- )t} \, , \quad
E _2 = e ^{-{\sqrt3} ( p_0 - p_- )t} \, , \quad
E _3 = e ^{-2{\sqrt3} ( p_0 + p_+ )t} \, , 		       &(3.17) \cr
{\om} _1 &= {{\rm i}\over {4{\sqrt 3}}} ( p_0 + p_+ - p_- ) \, ,  \quad
{\om} _1 = {{\rm i}\over {4{\sqrt 3}}} ( p_0 + p_+ + p_- ) \, , \quad
{\om} _1 = {-{\rm i}\over {4{\sqrt 3}}} p_+  \, ,                &(3.18) \cr}
$$
with $p_0$ given by (3.9).

	In the case of Bianchi II, the matrix $M$ appearing in (3.15,16)
must be of the form (3.3), and $(E _D , {\om} _D)$ can be expressed as
$$
\eqalignno {\! \! \! E _1 = e ^{-{\sqrt 3} ( p_0 + p_- )t} &, \,
E _2 = e ^{-{\sqrt 3} ( p_0 - p_- )t} 		  \, , \,
E _3 = {{2{\sqrt 3} \cosh ( 2{\sqrt 3} {\bar p_+} t )}\over {\bar p_+}}
e ^{-2{\sqrt 3}  p_0 t}                              , 	        &(3.19) \cr
&\quad {\om} _1 = {1\over 4{\sqrt 3}} \bigl ( F + {\rm i}
( p _0 - p _- - G ) \bigr )
	     				   \, \, \, , 		&(3.20a) \cr
	     {\om} _2  =& {1\over 4{\sqrt3}} \bigl ( F + {\rm i}
( p _0 + p _- - G ) \bigr )
	     				            \, , \, \, \,
	     {\om} _3  = {-1\over 4{\sqrt 3}} \bigl (F - {\rm i} G \bigr )
                                                    \, , 	&(3.20b) \cr
F =&{{\bar p_+}\over {\cosh ( 2{\sqrt 3} {\bar p_+} t )}}   \, , \quad
G = {\bar p_+} \tanh ( 2{\sqrt 3} {\bar p_+} t )
							\, , 	&(3.20c) \cr}
$$
where $( p _0, {\bar p} _+, p _- )$ must satisfy the restrictions (3.13).

	Even if the classical solutions that we have found correspond
to positive definite metrics, it is clear from (3.15) that one can always
reach degenerate metrics in the limits in which either $E _1, E _2, E _3$
or $\de M$ vanish. It is only in this sense that the degenerate solutions
are included in our analysis.

\noindent
{\bf IV. Bianchi Type II: Symplectic Structure of the Space of Solutions}
\medskip
\noindent
In this section, we will concentrate our attention on the type II Bianchi
model, studying the structure of the space of physical solutions.

	Let ${\vf} ^I$ be the set of left-invariant one-forms
for which the metric $g _{IJ}$ is diagonal. For Bianchi type II there
always exists a preferred one-form ${\vf} ^3$ selected by the symmetries
of the model, since the structure constants are given by
${C ^I} _{JK} = S^{IL} \e _{LJK}$, with $S ^{33}$ the only
non-vanishing component of $S^{IL}$. However,
the spacetime geometries remain obviously unaltered under the interchange
of ${\vf} ^1$ and ${\vf} ^2$. Two classical solutions which are related
by the interchange of indices $I=1$ and $I=2$ should then be identified as
the same physical solution.
\note{Nevertheless, one can neglect this identification
by considering ${\vf} ^1$ and ${\vf} ^2$ as two preferred one-forms.
The space of solutions will then correspond to a different theory, in
which the discussed symmetry is not present [9].}
{}From  the expressions (3.19,20), this interchange of indices
can be realized as a flip of sign in $p_-$. Therefore, we can restrict
ourselves only to non-negative parameters $p_- \geq 0$,
so that each physical solution is considered only once.

	There is still some redundancy left in the classical solutions of our
model. This redundancy comes from the fact that, if $A$ is a matrix that
satisfies
$$
\eqalignno { A ^t g _D A &= g _D 				&(4.1) \cr}
$$
for all diagonal metrics $g _D$, and such that $AM$ is of the form (3.3)
for every matrix $M$ of that form, the classical metrics (3.2) associated
to the collection of matrices $AM$ turn out to be identical [22].
The conditions
imposed on $A$ define a discrete group of four elements
$$
\eqalignno { \{ A _1 \equiv ( 1, 1, 1 ) \, , \, A _2 \equiv ( 1, -1, -1 ) \,
&, \, A _3 \equiv ( -1, -1, 1 ) \, , \, A _4 \equiv (-1, 1, -1) \}
						\, \, \,  , 	&(4.2) \cr}
$$
where $(a, b, c)$ denotes the ordered set of diagonal elements of $A$, and all
the non-diagonal elements are equal to zero. Using the invariance of the
physical solutions under multiplication of $M$ by $A_2$, we can choose
a positive determinant for the matrix ${\Me}$ appearing in (3.3). We will
thus restrict in the following to the case $\de {\Me} > 0$. Note that,
however, we have still to identify the classical solutions corresponding
to $M$ and $A_3M$, since multiplication by $A_3$ conserves the sign of
$\de {\Me}$. We will return to this point later in this section.

	Once we have determined the physically different classical solutions,
we proceed to show that the space of solutions is endowed with a symplectic
structure. We begin with the symplectic structure in the Ashtekar
formalism:
$$
\eqalignno { {\rm i} \Om &= d {A _I} ^a \w d {E ^I} _a \, \, \, . &(4.3) \cr}
$$
Substituting Eqs. (3.15,16) in this formula, we arrive at the expression
$$
\eqalignno { {\rm i} \Om &= d ( {\om} _I \de M ) \w \Bigl (
{dE_I\over E_I} + d (\de M) - {(dM M^{-1}) ^I} _I \Bigr )       \cr
			 &\z - {\om} _I \, \de M \,
{(dM M^{-1}) ^I} _J \w {(dM M^{-1}) ^J} _I 	\, \, \, , 	&(4.4) \cr}
$$
with ${(dM M^{-1}) ^I} _J = d{M ^I} _Q {(M^{-1}) ^Q} _J$ and
$d (\de M) = \de M \, \Tr (dM M^{-1})$. The complex orthogonal matrix
${\cal R} ^{aD}$ that is present in Eqs. (3.15,16) disappears completely
{}from the
symplectic form (4.4), as it represents only the gauge degrees of freedom
associated to the Gauss law constraints (2.9). On the other hand, taking
into account that the matrix
$M$ is of the form (3.3), with $\de {\Me} > 0$, we can always decompose
$M$ in the following product of matrices
$$
\eqalignno { M &\equiv M _D M _3 M _T R				\cr}$$
$$
\eqalignno{
&= \left ( \matrix { a & 0 &  0 \cr
		    0 & b &  0 \cr
		    0 & 0 & ab \cr} \right ) \,
  \left ( \matrix { 1 &              0 &  0 \cr
		    0 &              1 &  0 \cr
       {\twid M} ^3_1 & {\twid M} ^3_2 &  1 \cr} \right ) \,
  \left ( \matrix { 1 & 0 &  0 \cr
		    z & 1 &  0 \cr
                    0 & 0 &  1 \cr} \right ) \,
  \left ( \matrix { \cos \te  & \sin \te &  0 \cr
		  - \sin \te  & \cos \te &  0 \cr
                            0 &         0 &  1 \cr} \right )
						\, , 	&(4.5) \cr}
$$
with $a, b > 0$, ${\twid M ^3_1}, {\twid M ^3_2}, z \in \R$, and
$\te \in S^1$. The diagonal matrix $M_D$ can be absorbed into
the diagonal part of the classical solutions, $(E_D, {\om}_D)$,
by means of the redefinitions
$$
\eqalignno { {\what E} _1 = a b^2 E_1 \, , \quad
{\what E} _2 = a^2 b E_1 \, , \quad {\what E} _3 &= a b E_3 \, , \quad
{\what {\om} _D} = (ab) ^2 {\om} _D 			\, . 	&(4.6) \cr}
$$
Introducing then the notation ${\what M} = M _3 M _T R$, Eq. (4.4)
can be rewritten as
$$
\eqalignno { {\rm i} \Om &= d ({\what {\om} _I}) \w \Bigl (
d (\ln {\what E} _I) - {(d{\what M} {\what M} ^{-1}) ^I} _I \Bigr )
- {\what {\om}} _I {(d{\what M} {\what M} ^{-1}) ^I} _J
\w {(d{\what M} {\what M} ^{-1}) ^J} _I 		 \, , 	&(4.7) \cr}
$$
where we have employed that ${(dM_D {M^{-1}} _D) ^I} _J = \d ^I_J
({(d{M_D} {M^{-1}}_D) ^I} _I$.

	Let us define $S = M_T R$, so that ${\what M} = M _3 S$. It
is straightforward to compute that
$$
\eqalignno { {(d{\what M} {\what M} ^{-1}) ^I} _J &=
d({\twid M ^3_1}) \d ^I_3 \d ^1_J + d({\twid M ^3_2}) \d ^I_3 \d ^2_J
+ {(M _3) ^I} _Q {(dS S^{-1}) ^Q} _P {(M ^{-1}_3) ^P} _J
					 	 	\, . 	&(4.8) \cr}
$$
Then, using that ${(dS S^{-1}) ^I} _3 = 0$ and
$$
\eqalignno { {(M^{-1} _3) ^P} _I D_I {(M _3) ^P} _J &= D _I + D_I \d ^P_3
( {\twid M ^3_1} \d ^1_Q + {\twid M ^3_2} \d ^2_Q )
						 	&(4.9) \cr}
$$
for $D$ any diagonal matrix, we conclude that Eq. (4.7) is still valid
with the substitution of $S$ for ${\what M}$. In this way,
the matrix $(M_3)$ drops from the symplectic form $\Om$,
implying that ${\twid M ^3_1}$ and ${\twid M ^3_2}$
do not correspond to real degrees of freedom.

	In the parametrization chosen in (4.5), $(dS S^{-1})$ takes the
explicit expression
$$
\eqalignno { (dS S^{-1}) &=
\left (\matrix{- z d \te &   d \te & 0 \cr
d z - ( z ^2 + 1 ) d \te & z d \te & 0 \cr
                       0 &       0 & 0 \cr} \right )
						\, \, \, , 	&(4.10) \cr}
$$
{}from which it follows that
$$
\eqalignno { {\rm i} \Om &= d ({\what {\om} _D}) \w d (\ln {\what E} _D)
+ d\bigl ( ({\what \a} _1 - {\what \a} _2) z \bigr) \w d \te
						\, \, \, . 	&(4.11) \cr}
$$
Substituting Eqs. (4.6) in this formula, with $(E _D, {\om} _D)$
defined by means of (3.19,20) and (3.13), we arrive after some calculations
at a symplectic structure of the form
$$
\eqalignno { \Om &= d \Pi _+ \w dX + d \Pi _- \w dY + d \te \w dZ
						\, \, \, , 	&(4.12) \cr}
$$
where
$$
\eqalignno { \Pi _+ = (ab)^2 {\bar p_+} \, ,& \quad
\Pi _- = (ab)^2 p_-  \, , 					&(4.13) \cr}
$$
$$
\eqalignno { X = {3\over 4{\sqrt 3}} {\Pi _+\over\Pi _0} \ln (ab) \, ,&  \,
Y = {1\over 4{\sqrt 3}} \Bigl ( 3 {\Pi _-\over\Pi _0} \ln (ab) + \ln {a\over b}
\Bigr ) \, , \,
Z = {1\over 2{\sqrt 3}} {\Pi} _- z 	\, .                    &(4.14) \cr}
$$
Expression (4.12) provides us with the symplectic form for the space of
solutions to all physical constraints, \ie the reduced phase space.
Nevertheless, in obtaining this symplectic structure, we have implicitly
assumed
that the parameters $( p_+, p_-, p_0, t, a, b, z$, $\te$, ${\twid M} ^3_1,
{\twid M} ^3_2)$ are good coordinates in the space of physical solutions
to all but the scalar constraint; that is, that the transformation from
the triad and the spin connection to the given set of parameters is
analytic in the sector of the phase space
covered by these solutions. We will prove that this is
indeed the case in Sec. V.

	On the other hand, introducing the notation $\Pi _0 =
(ab) ^2 p _0$, the scalar constraint (3.6) implies that $(\Pi _+, \Pi _-)$
are a set of coordinates on the light-cone:
$$
\eqalignno { -{\Pi _0} ^2 +  {\Pi _+} ^2 + {\Pi _-} ^2 &= 0
						\, \, \, .	&(4.15) \cr}
$$
We point out that, in our model, ${\bar p_+}, a$ and $b$ are positive
quantities,
and that we have to restrict ourselves to positive $p_0$ and non-negative
$p_-$ in order to deal exclusively with different physical solutions.
With these restrictions, $(X, Y, Z)$ defined in (4.14) run still over
the whole real axis, for $z \in \R$. We can then interpret the two first
terms on the right hand side of (4.12) as the symplectic form of
the cotangent bundle over ${\cal L} ^+ _{(+,+)}$, the positive quadrant
of the future light-cone.

	In the last term of (4.12), the angle $\te$ belongs to $S ^1$.
We recall, however, that, from our previous discussion, there is
still some redundancy to be removed if we want to consider only physically
different classical solutions. This redundancy corresponds to the
identification of the matrix $M$ appearing in (4.5) with that obtained by
multiplication on the left by $A _3 = (-1, -1, 1)$. As a consequence,
the solutions associated with the parameters $(\te , {\twid M ^3_1},
{\twid M^3_2})$ and $(\te + \pi , - {\twid M ^3_1} , - {\twid M^3_2})$
are physically identical, and we may restrict our analysis
to the interval $\te \in [0, \pi)$. Note, nevertheless, that
${\twid M ^3_1}$ and ${\twid M^3_2}$ are not real degrees of freedom
(we can always go to the gauge in which ${\twid M ^3_1} = {\twid M^3_2}
= 0$), and that the coordinates $(\Pi _+, \Pi _-, X, Y, Z)$ remain
unaltered under the transformation
$(\te , {\twid M ^3_1}, {\twid M^3_2}) \to (\te + \pi , - {\twid M ^3_1} ,
- {\twid M^3_2})$. Therefore, the identification of physical
solutions under that transformation obliges us to identify also
the boundaries $\te =0$ and $\te=\pi$ of the reduced phase
space. In this way, the angle $\Theta = 2 \te$ turns out
belong to $S^1$, and the term $d \te \w d Z = d \Theta \w d ({Z\over 2})$
in (4.12) can be interpreted as the symplectic form of the cotangent bundle
over $S^1$. We thus conclude that the space of physical solutions in
Bianchi II presents the symplectic structure of the cotangent bundle
over the reduced configuration space ${\cal L} ^+_{(+,+)} \times S ^1$.

\noindent
{\bf V. Bianchi Type II: Analyticity of the Coordinatization of the Space\break
 of Solutions}
\medskip
\noindent
We want to prove that the parametrization employed to describe the
classical solutions for Bianchi type II (with $p_0$ regarded as
independent of $p_-$ and $p_+$) defines an analytic coordinatization in
the space of physical solutions to all but the scalar constraint,
in the sense that the transformation from the triad and the spin
connection to the chosen set of parameters is analytic in
the whole region of the phase space covered by the different
physical solutions that we have considered. We note first that
the transformation from the triad and the spin connection
to the three-metric and the extrinsic curvature is analytic in the
sector of the Ashtekar variables that corresponds to non-degenerate metrics.
It will suffice then to show that the matrix $M$, appearing in (3.2),
and the parameters $({\bar p_+}, p_-, p_0, t)$, contained in
$g_D$, depend analytically on $(g_{IJ} , K _{IJ})$ in the region
defined by $\de {\Me}, p_+, p_- > 0$, which contains all the different
physical solutions.

	Let us introduce a matrix ${\bar M}$ of positive determinant
(and thus invertible) such that it satisfies the conditions
$$
\eqalignno { {\bigl ( ( {\bar M} ^{-1} ) ^t \bigr ) _I} ^P g _{PQ}
{( {\bar M} ^{-1} ) ^Q} _J &= \d _{IJ} + ( {\bar g} _3 - 1 )
\d ^3_I \d ^3_J					\, \, \, , 	&(5.1) \cr
{\bigl ( ( {\bar M} ^{-1} ) ^t \bigr ) _I} ^P K _{PQ}
{( {\bar M} ^{-1} ) ^Q} _J &= {\l} {\d} _{IJ}    \, \, \, . 	&(5.2) \cr}
$$
It is clear from our previous analysis of the type II Bianchi model that
one solution to Eqs. (5.1,2) is provided by
$$
\eqalignno { {\bar M} &= \Bigl (
(g _1) ^{\half} , (g _2) ^{\half} , (g _1 g _2 ) ^{\half}
\Bigr ) \, M     				\, \, \, , 	&(5.3) \cr}
$$
where we have used a similar notation to that displayed in (4.2),
$M$ is the matrix appearing in (3.2,3), and $(g_1,g_2)$ are given by
(3.11) in our parametrization. From Eqs. (3.11,12) and
(3.14), one can compute also the explicit expressions of ${\bar g_3}$ and
${\l}_I$ for the solution (5.3):
$$
\eqalignno { {\bar g_3} &= 4 {\l} ^2 \Bigl (
{{\bar p_+}\over {\cosh (2{\sqrt 3}{\bar p_+}t)}} \Bigr) ^2 \, , \quad
{\l} _1 = {\l} \bigl ( - p_0 + p_- + {\bar p_+} \tanh (2{\sqrt 3}{\bar p_+}t)
\bigr )  						\, , 	&(5.4a) \cr
{\l} _2 &= {\l} \bigl ( - p_0 - p_- + {\bar p_+} \tanh (2{\sqrt 3}{\bar p_+}t)
\bigr ) 						 \, , \quad
{\l} _3 = - {\l} {\bar g_3} {\bar p_+} \tanh (2{\sqrt 3}{\bar p_+}t)
							 \, , 	&(5.4b) \cr
&\qquad \qquad \qquad
{\l} = {e^{2{\sqrt 3}p_0t}\over {4{\sqrt 3}}} \Bigl (
{{\bar p_+}\over {2{\sqrt 3} \cosh (2{\sqrt 3}{\bar p_+}t)}} \Bigr ) ^{\half}
						\, \, \, . 	&(5.4c) \cr}
$$

	Since $g _1$ and $g _2$, given by (3.11), are strictly positive,
Eq. (5.3) defines $M$ analy\-ti\-cally in terms of
$({\bar p_+}, p_-, p_0, t, {\bar M})$.
All we have to prove then is that $({\bar p_+}, p_-, p_0, t, {\bar M})$
can be obtained  analytically from $( g_{IJ} , K _{IJ} )$ in the region
of the phase space that we are considering. On the other hand, it is easy
to check that the matrix $\bar M$ in (5.3) is of the form (3.3),
provided that $M$ is of this form. Moreover, adopting a parallel notation
to that introduced in Eq. (3.3), one can see that
${\de} {\bar {\Me}} > 0$ if ${\de} {\Me} > 0$. We will thus concentrate
in the rest of this section on matrices ${\bar M}$ of the type (3.3),
for which ${\de} {\bar {\Me}} > 0$, and such that they verify Eqs. (5.1,2).

	It is convenient to employ the following decomposition for the
matrix $\bar M$:
$$
\eqalignno { &{\bar M} \equiv {\bar R} {\bar M_D} {\bar M _T}{\bar M _3}
							\, , 	&(5.5) \cr}
$$
with ${\bar R}, {\bar M_D}, {\bar M _T}$, and ${\bar M _3}$ parametrized
as their respective counterparts in Eq. (4.5) with the substitutions
$(a , b , {\twid M ^3_1} , {\twid M ^3_2} , z , \te ) \to
( {\bar a} , {\bar b} , {\bar M ^3_1} , {\bar M ^3_2} , {\bar z} ,
{\bar \te} )$, so that
${\bar a}, {\bar b} > 0$, ${\bar z}, {\bar M} ^3_1 ,
{\bar M} ^3_2 \in \R$ and ${\bar \te} \in S^1$. Let us show then that all
the elements of these matrices
present, through Eqs. (5.1,2), an analytic dependence on
$( g _{IJ} , K _{IJ} )$ in the region of physical interest.

	We first determine ${\bar M _3}$ by the conditions
$$
\eqalignno { {\twid g} _{13} &= {\twid g} _{23} = 0 \, , \, \,
{\rm with} \quad {\twid g} _{IJ} =
{\bigl ( ( {\bar M _3} ^{-1} ) ^t \bigr ) _I} ^P g _{PQ}
{( {\bar M _3} ^{-1} ) ^Q} _J 		\, ,			&(5.6) \cr}
$$
{}from which one obtains that ${\bar M} ^3_1 = g_{13}/g_{33}$ and
${\bar M} ^3_2 = g_{12}/g_{33}$. ${\bar M_3}$ is thus analytic
in $g_{IJ}$, since $g_{IJ}$ is a positive definite metric, and
therefore $g_{33}>0$.
Let us define then the two-by-two matrix ${\twid h}$ constructed with the two
first rows and columns of ${\twid g}$, and, si-\break milarly, the matrix
${\twid {\cal K}}$ obtained from the extrinsic curvature ${\twid K} =
({\bar M _3} ^{-1}) ^t K ({\bar M _3} ^{-1})$.
\note{From our discussion of the type II Bianchi model in Sec. III,
it follows that, given the form of ${\twid g}$, ${\twid K}$ must satisfy
${\twid K} _{13} = {\twid K} _{23} = 0$ if $({\twid g} _{IJ} ,
{\twid K} _{IJ})$ corresponds to a classical solution to all but the
Hamiltonian constraint.}
We can use the degrees of freedom ${\bar a}$, ${\bar b}$, and ${\bar z}$ in
${\bar M}$ to diagonalize ${\twid h}$ to the identity
$$
\eqalignno { ({\bar {\Me} _D} ^{-1} ) ^t ({\bar {\Me} _T} ^{-1} ) ^t \,
{\twid h} \, ({\bar {\Me} _T} ^{-1} ) ({\bar {\Me} _D} ^{-1} ) &= {I\!\!\!I}
	     					\, \, \, , 	&(5.7) \cr}
$$
where ${\bar {\Me} _D}$ and ${\bar {\Me} _T}$ are the two-by-two matrices
constructed from ${\bar M_D}$ and ${\bar M_T}$ by the procedure explained
above. Eq. (5.7) fixes ${\bar a}$, ${\bar b}$, and ${\bar z}$ uniquely,
for ${\bar a}$ and ${\bar b}$ strictly positive,
$$
\eqalignno { {\bar a} = \Bigl ( {{\de {\twid h}}\over {{\twid h}_{22}}}
\Bigr ) ^{\half} \, , \quad {\bar b} &= ({\twid h} _{22}) ^{\half} \, , \quad
{\bar z} = {{\twid h} _{12}\over {{\twid h} _{22}}} 	\, .	&(5.8) \cr}
$$
{}From the expression (5.8) we conclude that
${\bar a}$, ${\bar b}$, and ${\bar z}$ are analytic in ${\twid h}$
(and then in $g$) for ${\twid h}$ positive definite.

	We can now diagonalize ${\what {\cal K}} =
({\bar {\Me} _D} ^{-1} ) ^t ({\bar {\Me} _T} ^{-1} ) ^t
{\twid {\cal K}} ({\bar {\Me} _T} ^{-1} ) ({\bar {\Me} _D} ^{-1} )$
by the $SO(2)$ transformation given by the matrix ${\bar {\cal R}}$
contained in ${\bar R}$. This transformation leaves invariant
the value ${\what h} = {I\!\!\!I}$ reached in (5.7).
We are thus left with the eigenvalue problem
$$
\eqalignno { F _{IJ} &\equiv {{\bar {\cal R}} _I} ^{\z P}
{\what {\cal K}} _{PQ} {({\bar {\cal R}} ^t) ^Q} _J - {\l} _I \d _{IJ} = 0
						\, \, \, . 	&(5.9) \cr}
$$
The element $F _{12}$ of the system of equations (5.9) defines $\te$ as
an implicit function of ${\twid {\cal K}}$.
Taking into account that, from our previous analysis, ${\twid {\cal K}}$
is analytic in $( g_{IJ} , K _{IJ} )$,
all we have to prove is that the implicit dependence of $\te$ on
${\twid {\cal K}}$, imposed by Eq. (5.9), is analytic.
Let us suppose that $( {\te} ^0 , {{\twid {\cal K}} ^0} _{IJ} )$
is a particular solution to the equation $F_{12}=0$\break
that determines $\te$
as a function of ${\twid {\cal K}}$. Then, $\te ({\twid {\cal K}} ^0)
= {\te} ^0$ defines an analytic germ [23] around ${\twid {\cal K}} ^0$
which can be continued analytically as long as $\partial _{\te} F _{12}
 \bigl ( {\te} ({\twid {\cal K}}) , {\twid {\cal K}} \bigr ) \not= 0$\break
[23,24] (in particular, if ${\te} ^0$ and ${{\twid {\cal K}} ^0} _{IJ}$
are real and ${\te} ({\twid {\cal K}})$ can be continued analytically
around ${{\twid {\cal K}} ^0} _{IJ}$, it is possible to show that
${\te} ({\twid {\cal K}})$ remains real for ${\twid {\cal K}} \in \R$).
After a simple computation, one arrives at the identity
$$
\eqalignno { \partial _{\te} F _{12} &= {\l} _2 - {\l} _1
						\, \, \, , 	&(5.10) \cr}
$$
so that the obtained solution ${\te} ({\twid {\cal K}})$ depends analytically
on ${\twid {\cal K}}$ as far as ${\l} _1 \not= {\l} _2$.
\note{Eq. (5.9) can be interpreted as an eigenvalue problem. Each row
of the matrix ${\bar {\cal R}}$ can be identified as an unit eigenvector
of the matrix ${\twid {\cal K}}$. When ${\l} _1 = {\l} _2$, the eigenvalue
problem is degenerate: any unit vector is an eigenvector of
${\twid {\cal K}}$. As a consequence, ${\bar {\cal R}}$ is ill-defined in that
case.}
Note that, in our parametrization, ${\l}_1={\l}_2$ only if $p_-=0$
(see Eqs. (5.4)). Since we are restricting our attention to the sector of
positive definite metrics and extrinsic curvatures for which
$p_- \geq 0$, we conclude that ${\te} ({\twid {\cal K}})$ is analytic
in that region, except at the boundary $p_0 = 0$.
\note{It is clear then that, without a restriction of this type, $p_-$
would have not been a good coordinate in the phase space of the full
type II Bianchi model.}
With this caveat, the matrix ${\bar M}$ turns out then to be analytic
in the sector of metrics and extrinsic curvatures associated to
the physical solutions analyzed in Sec. IV.

	Employing this result, Eqs. (5.1,2) provide us with ${\bar g_3}$ and
${\l} _I \, (I= 1, 2, 3)$ as analytic functions of
$(g _{IJ} , K _{IJ})$, for $({\bar M} ^{-1})$ is always well-defined, and is
analytic if ${\bar M}$ is analytic. We can use now the explicit
expressions (5.4) to determine $({\bar p_+} , p_-, p _0 , t)$
as analytic functions of ${\bar g_3}$ and ${\l} _I$, and, therefore,
of $(g _{IJ} , K _{IJ})$. The relations obtained are
$$
\eqalignno { p _- &= {\bar p _+} {F _1\over F _3}  \, , \quad
p _0 = {\bar p _+} {F _2\over F _3} 			\, ,	&(5.11a) \cr
t &= {1\over {2{\sqrt 3}{\bar p _+}}} {\cosh } ^{-1}
\bigl ( {F_3\over F_4} \bigr )		 		\, , 	&(5.11b) \cr
{\bar p_+} &= 2{\sqrt 3} F _3 \Bigl ( {4\over F _4}
e ^{-2 {F_2\over F_3} {\cosh } ^{-1} ( {F_3\over F_4} ) } \Bigr ) ^{1\over 3}
							\, , 	&(5.11c) \cr}
$$
where
$$
\eqalignno { F _1 &= {\half} ( {\l} _1 - {\l} _2 ) \, , \quad
F _2 = - {{\l} _3\over {\bar g_3}} - {\half} ( {\l} _1 + {\l} _2 )
						\, , 	&(5.12a) \cr
F _3 &= \Bigl ( {{\bar g_3}\over 4}
+ \Bigl ( {{\l} _3\over {\bar g_3}} \Bigl ) ^2  \Bigr ) ^{\half}
\, , \quad
F _4 = {({\bar g_3}) ^{\half}\over 2} 	      		\, .	&(5.12b) \cr}
$$
For real positive definite metrics (so that ${\bar g_3} >0$) and
real extrinsic curvatures, all the functions $F_n$ $($with $n=1,...,4)$
are analytic in $({\bar g_3}, {\l} _I)$, and $F_3$ and $F_4$ turn out to be
positive. Then, $({\bar p_+}, p_-,p_0,t)$, given
by (5.11), result to be analytic in $(g_{IJ},K_{IJ})$.

{}From our previous discussion, it follows that the matrix $M$,
defined by means of (5.3), is analytic in $(g _{IJ} , K _{IJ})$ in the
sector of solutions to all but the Hamiltonian constraint that we are
studying. The only point that remains to be proved is that the specific
parametrization (4.5), employed for $M$ in Sec. IV, is analytic with
respect to its dependence on the elements of $M$. Using this
parametrization, and the notation in (3.3), we have that
$$
\eqalignno { {\Me} =&
\left ( \matrix { {M ^1} _1 & {M ^1} _2 \cr
	   	  {M ^2} _1 & {M ^2} _2 \cr} \right ) =
\left ( \matrix {   a & 0 \cr
		  b z & b \cr} \right ) \,
\left ( \matrix { \cos \te & \sin \te \cr
		- \sin \te & \cos \te \cr} \right )
 						     \, , 	&(5.13) \cr
             &M ^3_1 = ab \Bigl ( {\twid M} ^3_1 \cos {\te} + {\twid M} ^3_2
(z \cos {\te} - \sin {\te} ) \Bigr ) 		\, \, \, , 	&(5.14a) \cr
             &M ^3_1 = ab \Bigl ( {\twid M} ^3_1 \sin {\te} + {\twid M} ^3_2
(z \sin {\te} + \cos {\te} ) \Bigr ) 		\, \, \, , 	&(5.14b) \cr}
$$
where $a$ and $b$ are strictly positive, and ${\det {\Me}} > 0$
by construction. From Eq. (5.13), one can easily
arrive at the explicit expressions of $a , b ,$ and $z$ in terms of
${\Me} {\Me} ^t$. These expressions are analytic in the elements of
${\Me}$ for ${\de } {\Me} > 0$, and allow us to define $a$ and $b$ as strictly
positive functions: $a ({\Me}), b ({\Me}) > 0$.
Eq. (5.13) provides us also with the relations $\cos \te = M^1_1 / a(\Me)$
and $\sin \te = M^1_2 / a(\Me)$, from which we conclude that $\cos \te$
and $\sin \te$ are analytic in ${\Me}$, as $a ({\Me}) > 0$.
\note{Indeed, these relations define the angle $\te$ analytically,
because the Jacobian matrix of $(\cos \te , \sin \te)$ with
respect to $\te$ has always a rank equal to the unity.}
Substituting then $a , b, z, \cos \te$ and $\sin \te$ as functions of
$\Me$, the equations in (5.14) determine ${{\twid M ^3}_1}$ and
${{\twid M ^3}_2}$ as two implicit functions of the elements of the matrix
$M : ({{\twid M} ^3}_1 (M), {{\twid M} ^3}_2 (M))$.
The Jacobian of these equations
with respect to ${{\twid M} ^3}_1$ and ${{\twid M} ^3}_2$ can be computed
straightforwardly to be equal to $ a ^2 ({\Me}) b ^2 ({\Me})$. Therefore,
given a particular analytic germ for $({{\twid M} ^3}_1,{{\twid M} ^3}_2)$,
we can always continue it analytically to the whole range of matrices
$M$ with ${\de} {\Me} > 0$ (so that $a ({\Me})$ and $b ({\Me})$ are
strictly positive) [23,24].

	In conclusion, we have proved that the parameters
$( p_+, p_-, p_0, t, a, b, z, \te, {\twid M} ^3_1, {\twid M} ^3_2)$
are good coordinates in the space of non-degenerate real solutions
that satisfy all but the Hamiltonian constraint, when restricted to
the region
$$
\eqalignno { {\bar p_+}, p_-, a, b > 0 \, , \quad \te &\in S ^1 \, , \quad
p_0, t, z, {\twid M} ^3_1, {\twid M} ^3_2 \in \R 	\, . 	&(5.15) \cr}
$$
If we want to go to the reduced phase space, we have to impose in addition
the constraint (3.6), and restrict the range of $p_0$ to
the positive real axis. As a consequence, the symplectic
structure obtained in Sec. IV results to be analytic everywhere
in the cotangent bundle over the reduced configuration space
${\cal L} ^+_{(+,+)} \times S ^1$, with its boundary excluded.

\noindent
{\bf VI. Type I Bianchi Model}
\medskip
\noindent
Let us proceed now to generalize the study of Secs. IV and V
to the case
of the type I Bianchi model. We begin by analyzing the sector of
classical solutions that corresponds to different spacetime geometries.

	In the set of left-invariant one-forms ${\vf} ^I$
in which the metric is diagonal, the spacetime geometries
are invariant under permutations of all ${\vf} ^I$,
because the structure constants vanish identically in this model,
so that there is no preferred one-form. Therefore,
the classical diagonal solutions that are related under any interchange of
the indices $I= 1, 2, 3$ must be identified. From expressions
(3.17,18), each of the planes $\Pi _1 \equiv p_- = 0, \, \Pi _2 \equiv p_0 +2
p_+ - p_- = 0$ and $\Pi _3 \equiv p _0 + 2p_+ + p_- = 0$ divide the space
$(p_0,p_+,p_-) \in {\R} ^3$ into
two regions which can be interchanged under the
respective permutations of indices: $1\leftrightarrow 2$, $1\leftrightarrow 3$
and $2\leftrightarrow 3$. If we take into account also the constraint
(3.9), which implies that $(p_0,p_+,p_-)$ lie in the future light-cone,
the requirement of considering only different classical solutions may be
implemented by the following restrictions in the ranges of our parameters:
$$
\eqalignno { p _+ \geq 0 \, , \quad p_- \geq 0 \, &\quad
p _0 = \sqrt {{p _+} ^2 + {p _-} ^2} \geq 0 	\, , 	&(6.1) \cr}
$$
\ie $(p_0,p_+,p_-) \in {\cal L} ^+_{(+,+)}$.

	Parallel to the situation in Bianchi type II, we have to identify
also the solution (3.15,16) for a given matrix $M\in GL(3,\R)$ with all
other solutions obtained from matrices of the form $AM$, where $A$
is any orthogonal matrix that
commutes with the diagonal subgroup of $GL(3, \R)$ (see Eq. (4.1)). We can then
fix the determinant of $M$ to be strictly positive, because either $M$ or
$-M$ has a positive determinant, and $A=(-1, -1, -1)$ is orthogonal and
commutes with the diagonal subgroup. There is, however, some redundancy left,
because there exist matrices $A$ which conserve the sign of the determinant
of $M$. These matrices are given by the discrete group of four elements
defined in (4.2) [22]. We will discuss the corresponding identification of
classical solutions later in this section. Finally, we point out that,
given the explicit expression (3.7) for the diagonal metric in
Bianchi type I, it is always possible to absorb the determinant of $M$
by a redefinition of the origin of the time coordinate t.
We will thus restrict hereafter to matrices $M \in SL (3, \R)$.

	Let us introduce the following parametrization for the matrices
$M \in SL (3, \R)$:
$$
\eqalignno { M & \equiv M_D M_T R =
\left ( \matrix { a & 0 &           0 \cr
                  0 & b &           0 \cr
                  0 & 0 & {1\over ab} \cr} \right ) \,
\left ( \matrix { 1 & 0 & 0 \cr
                  x & 1 & 0 \cr
                  y & z & 1 \cr} \right ) \, \Bigl ( R \Bigr )
				                	\, , 	&(6.2) \cr}
$$
where $a$ and $b$ are strictly positive, $x, y$ and $z$ are real and
$R \in SO(3)$. Similar to the notation for Bianchi type II, we will
call
$$
\eqalignno { {\what E} _1 = {E _1\over a} \, , \quad
{\what E} _2 &= {E _2\over b} \, , \quad {\what E} _3 = ab E _3
							\, . 	&(6.3) \cr}
$$
Then, it is straightforward to generalize the discussion presented in
Sec. IV to arrive at a symplectic structure of the form (4.7)
for the space of solutions in Bianchi type I,\break
with
${\what M} = M _T R$ and ${\what {\om}} _I = {\om} _I$ (note that now
$\de M = 1$).

	The symplectic form (4.7) can be simplified as follows. Decompose
first the matrix ${\what M}$ as ${\what M} = M _T R$, and define $T ^{IJ} =
d R ^{IP} R ^{JP}$. From the expression of $M_T$ in (6.2), one can
explicitly check that
$$
\eqalignno { \sum _P d {(M_T) ^I} _P {({M _T} ^{-1}) ^P} _I &= 0
							 	&(6.4) \cr}
$$
for every $I = 1, 2, 3$. On the other hand, the matrix of one-forms
$T ^{IJ}$ turns out to be antisymmetric, since $R^{IP} R^{JP} = \d ^{IJ}$,
and so $dR^{IP} R^{JP} = - dR ^{JP} R^{IP}$. Using this fact, it is possible
to prove that
$$
\eqalignno { dT ^{IJ} &= T ^{IP} \w T ^{PJ} 	\, \, \, .	&(6.5) \cr}
$$
Conveniently rearranging the different terms in (4.7), we arrive
at the conclusion
$$
\eqalignno { {\rm i} \Om &= d {\what {\om} _I} \w d \ln ({\what E} _I)
- d ( L _{IJ} T^{IJ} ) 	  			\, \, \, ,	&(6.6) \cr
L _{IJ} &= ( {M_T} ^{-1} ) _{[I|P} {\om} _P {( M_T ) ^P} _{|J]}
						\, \, \, , 	&(6.7) \cr}
$$
where the indices $I,J$ are raised and lowered with the metric ${\eta} _{IJ}
= (1, 1, 1)$ and $[ \z | \z]$ denotes antisymmetrization. The first term on
the right hand side of Eq. (6.6) can be calculated from Eqs.
(6.3), (3.17,18) and (3.9). The result is
$$
\eqalignno { d {\om} _I \w d \ln ({\what E} _I) &= {{\rm i}\over 4{\sqrt3}}
( dp _+ \w dX + dp _- \w dY ) 			\, \, \, , 	&(6.8) \cr}
$$
with
$$
\eqalignno { X = - \Bigl ( {p_+\over p_0} + 2 \Bigr ) \ln (ab) \, , &\quad
Y= \ln \Bigl ( {a\over b} \Bigr ) - {p_-\over p_0} \ln (ab)
							\, .	&(6.9) \cr}
$$
The one-form $\Tr ( - L T )$ that appears in (6.6) can be interpreted as
the pre-symplectic structure in $SO(3)$. From the parametrizations
for ${\om} _I$ and $M_T$ given by (3.18)\break
and (6.2), one can explicitly compute
the matrix $L_{IJ}$. Introducing the notation \break
$L_{IJ} = \e _{IJK} l^K$, we find that
$$
\eqalignno { & \qquad \qquad l _1 = (p_0 + 2 p_+ + p_-) z
					\, \, \, , 	&(6.10a)\cr
	     l _2 &= - \Bigl ( 2 p_- (xz-y) + (p_0 + 2 p_+ + p_-) y \Bigr )
\, , \, \,  l _3 = -2 p_- x 		   \, . 	&(6.10b) \cr}
$$
Let us parametrize now the matrices $R \in SO(3)$ in terms of the Euler
angles $( \a , \b , \te)$
$$
\eqalignno { R ( \a , \b , \te ) &= R_{(1)} (\a) R_{(3)} (\b) R_{(1)} (\te)
						\, \, \, , 	&(6.11) \cr}
$$
where $R_{(I)} (\a)$ is a rotation of an angle $\a$ around the axis defined by
the direction $I$,\break
and $\a , \te \in S^1 \, , \b \in [0,\pi]$. This
parametrization is unique and well-defined for all matrices $R \in SO(3)$,
except at $\b = 0$ and $\b=\pi$. After a short calculation using Eqs.\break
(6.10,11), we conclude that the pre-symplectic structure $\Tr ( - L T )$
can be written as
$$
\eqalignno { L _{IJ} T^{IJ} &= {{\rm i}\over 4{\sqrt 3}} ( F _1 d{\a} +
F _2 d{\b} + F _3 d{\te} ) 			\, \, \, , 	& (6.12) \cr}
$$
with
$$
\eqalignno { F _1 &= l _1 			\, , \quad
	     F _2  = l _2 \sin \a + l _3 \cos \a
                                                \, \, \, , 	&(6.13a) \cr
F _3 &= - l _3 \sin \a \sin \b + l _2 \cos \a \sin \b + l _1 \cos \b
						\, \, \, . 	& (6.13b) \cr}
$$
Therefore, the symplectic form (6.7) has the expression
$$
\eqalignno { \Om &= {1\over 4{\sqrt 3}}	( d p _+ \w dX + d p _- \w dY +
d \a \w d F_1 + d \b \w d F _2 + d \te \w dF_3 ) 	\, . 	&(6.14) \cr}
$$
For $a$ and $b$ strictly positive and $x, y, z$ real, the variables
$X,Y,F_1,F_2$ and $F_3$ run over the whole real axis $\R$. As
it stands, the symplectic structure (6.14) might be interpreted as that
corresponding to the cotangent bundle over ${\cal L} ^+ _{(+,+)} \times
SO(3)$. Nevertheless, we have still to identify the classical solutions
obtained from all matrices of the form $AM$, with $M \in SL(3, \R)$
and $A$ any matrix in the discrete group (4.2). In the parametrizations
(6.2) and (6.11), the change of $M$ to $A_2 M$ can be realized as the
transformation $(\a , x, y ) \to ( \a + \pi, -x, -y)$, which leaves invariant
all the variables appearing in (6.14) except $\a$
(see Eqs. (6.10) and (6.13)). In order to consider only
different physical solutions, we can restrict
ourselves to the range $\a \in [ 0 , \pi )$ [22]. The boundaries $\a = 0$ and
$\a = \pi$ must be identified in the reduced phase space, since, from our
previous discussion, the classical solutions for $\a = 0$ and $\a = \pi$
are physically identical. In this way, ${\twid \a} = 2 \a$ belongs to $S^1$,
and we can
interpret the symplectic form $d \a \w d F_1 + d \b \w d F _2$ as that of the
cotangent bundle over the two-sphere, $S^2$, parametrized by the angles
$\b$ and ${\twid \a}$. The singularities $\b = 0$ and $\b = \pi$ of the
parametrization (6.11) then correspond to the poles of this two-sphere.

	Let us now study the identification of the matrices $M$ and
$A _3M$. The interchange of these two matrices is performed by the
transformation $( \a , \b , \te , y, z) \to ( \pi - \a $, \break
$\pi - \b , \te + \pi , -y, -z)$, where, we recall, $\a \in [ 0 , \pi )$.
We could then impose the restriction $\te \in [ 0 , \pi )$,
so that each physical solution is considered only once [22].
Note that, under the above transformation, $p_+ , p_-, X, Y$
and $F _3$ remain invariant, while $F _1$ and $F _2$ flip their sign.
We should thus identify the points
$(\a , \b , \te = 0 , F _1 , F _2 , F _3 )$ and
$ ( \pi - \a , \pi - \b , \te = \pi , - F _1, - F _2, F _3 )$
at the boundaries $\te = 0$ and $\te = \pi$ of the reduced
phase space. Nevertheless, we will adopt a different approach
for the quantization of the model, leaving $\te$ to run
over the whole of $S ^1$ and imposing restrictions on the space of
physical states associated to the identification of points
$$
\eqalignno { (\a , \b , \te , F _1 , F _2 , F _3 ) \quad  & {\rm and} \quad
 ( \pi - \a , \pi - \b , \te + \pi , - F _1, - F _2, F _3 ) \, . &(6.15) \cr}
$$
We will return to this issue in the next section.

	Finally, the matrices $M$ and $A_4M$ are interchanged
through the transformation
$(\sin \a , \b , \te , x , z )
\to (-\sin \a , \pi - \b , \te + \pi , -x, -z )$,
which leaves invariant $p_+ ,p_- , X, Y,$ and $F_3$, and reverses the signs of
$F_1$ and $F_2$. For $\a \in [ 0 , \pi ) , \sin \a$ is non-negative, and the
only possible redundancy left is at $\a = 0$. However, we notice that the
corresponding identification of points $(\a = 0 , \b , \te , F _1 , F _2 ,
F _3 )$ and $( \a = 0  , \pi - \b $, \break
$\te + \pi , - F _1, - F _2, F _3 )$
has already been taken into account by identifying the points in (6.15),
and $\a$ with $\pi + \a$.

	In conclusion, the space of physical solutions in the type I
Bianchi model has the symplectic structure of the cotangent bundle
over the reduced configuration space ${\cal L} ^+ _{(+,+)} \times S ^2
\times S ^1$. To consider only different physical solutions we still
must impose the identification of points described in (6.15). We will
implement this condition in the quantum version of the model as a
restriction on the physical states.

	We have been assuming so far that the parametrization used for the
classical solutions corresponds to a good set of coordinates in the phase
space. In fact, employing parallel arguments to those presented in Sec. V
for the type II Bianchi model it is possible to show that the transformation
{}from the triad and the spin connection to the chosen
set of parameters is analytic in the considered
physical solutions to all but the scalar constraint
(except at $\b = 0 , \pi$). We will now briefly discuss the main lines
of this proof.

	Let $(g _{IJ}, K_{IJ})$ be the metric and extrinsic curvature for
the type I Bianchi model, which are analytic in the triad and the spin
connection for $g_{IJ}$ a non-degenerate metric. We first introduce
the matrix ${{\bar M} ^I} _J = (g _I ) ^{\half } {M ^I} _J$,
where ${M ^I}_J \in SL (3, \R)$ is the matrix which appears in (3.2)
and leads to a diagonal form for the metric $g_{IJ}$, and
$g_I$ are the components of the corresponding diagonal metric.
The matrix ${\bar M}$ satisfies the relations
$$
\eqalignno { \de {\bar M} > 0 \, , \quad  {\bigl ( ({\bar M} ^{-1} ) ^t
\bigr ) _I} ^P &g _{PQ} {({\bar M} ^{-1} ) ^Q} _I = \d _{IJ}
					   		\, , 	&(6.16) \cr
            {\bigl ( ({\bar M} ^{-1} ) ^t \bigr ) _I} ^P K _{PQ}&
{({\bar M} ^{-1} ) ^Q} _I = {\l} _I \d _{IJ}		\, , 	&(6.17) \cr}
$$
with
$$
\eqalignno { {\l} _I = {{\dot g} _I \over {2N g_I}} \, , &\quad
\de {\bar M} = ( g _1 g _2 g _3 ) ^{\half} 		\, . 	&(6.18) \cr}
$$
One can use Eqs. (6.16,17) to determine ${\bar M}$ in terms of
$g _{IJ}$ and $K _{IJ}$. It is convenient to adopt
the decomposition ${\bar M} = {\bar R} {\bar M_T}$, with
${\bar R} \in SO(3)$, and ${\bar M_T}$ a lower-triangular matrix
that includes diagonal elements different from the unity. Eq. (6.16)
fixes ${\bar M_T}$ analytically as a function of $g_{IJ}$.
On the other hand, the non-diagonal
components of the system of equations (6.17) provide us with the matrix
${\bar R}$ as an analytic function of $(g_{IJ}, K_{IJ})$, as far as
${\l} _I \not= {\l} _J$ for all different $I,J = 1,2,3$. From expressions
(6.18) and Eqs. (3.7,8), one can easily check that this is indeed the case
if $p_+ , p_- $, \break
$( p _0 + 2 p _+ - p _- ) > 0$. Substituting the resulting matrix
${\bar M}$,
the diagonal components of the system (6.17) determine ${\l} _I$ analytically
in terms of $( g _{IJ} , K _{IJ} )$. One can then identify ${\l} _I$
and $\de {\bar M}$ with their explicit expressions in our parametrization
(through Eqs. (6.18)), and subsequently compute the parameters
$p_0 , p_+ , p_-$ and $t$ as analytic
functions of $(g _{IJ} , K _{IJ})$. Inserting these functions into the
expressions (3.7) for $g _I$, we obtain the matrix ${M ^I} _J =
(g_I) ^{- \half} {{\bar M} ^I} _J$, which turns out to depend analytically
on $(g _{IJ} , K _{IJ})$. Finally, the parametrization used for ${M ^I} _J$
in (6.2) and (6.11) is well-defined and analytic, except at $\b = 0$ and
$\b = \pi$, points that can be interpreted as the poles of the two-sphere
coordinatized by ${\twid \a} = 2 \a$ and $\b$. We consider then the
non-analyticity at $\b =0$ and $\b = \pi$ as a failure in the coordinatization
of $S ^2$. In order to go to the reduced phase space, one has only
to impose the scalar constraint (3.6) on $p_0$, $p_+$ and $p_-$,
and restrict oneself exclusively to positive $p_0$ (then,
$(p _0 + 2 p _+ - p _-) > 0$ is automatically satisfied).

\noindent
{\bf VII. Bianchi types I and II: Quantization}
\medskip
\noindent
Once we have identified the symplectic structures and the reduced
configuration spaces for Bianchi I and II, we turn to the task of the
canonical quantization of these models. Let us start with Bianchi type II.
In this case the reduced phase space is the cotangent bundle over
${\cal L} ^+ _{(+,+)} \times S ^1$. Keeping the notation
introduced in Sec. IV, we parametrize $S^1$ by the angle $\Te$, while
${\cal L} ^+ _{(+,+)}$ is defined by the constraint $\Pi _0 =
\sqrt {{\Pi _+} ^2 + {\Pi _-} ^2}$, with $\Pi _0, \Pi _+, \Pi _- \in
\R ^+$. A natural set of elementary variables in this reduced
configuration space is then provided by $(\Pi _+, \Pi _- ,
c \equiv \cos \Te , s \equiv \sin \Te)$. As generalized momentum
variables we choose [9]
$$
\eqalignno { L _+ &= \Pi _+ U + {\Pi _+ ^2 \over \Pi _0} T
                                                              \, , &(7.1) \cr
L _- &= \Pi _- V + {\Pi _- ^2 \over \Pi _0} T
                                                              \, , &(7.2) \cr
L _{\Te} &= Z
                                                              \, , &(7.3) \cr}
$$
where the momenta $(T , U , V , Z)$ are canonically conjugate to
$(\Pi _0, \Pi _+, \Pi _- , \Te )$. Our set of reduced phase space variables
commute
with the constraint (4.15), with respect to the Poisson-brackets structure.
On the other hand, the only non-vanishing
Poisson-brackets among the configuration and momentum variables are:
$$
\eqalignno { \{ \Pi _+ , L_+ \} &= \Pi _+  		\, , 	&(7.4) \cr
	     \{ \Pi _- , L_- \} &= \Pi _-		\, , 	&(7.5) \cr
            \{ c , L _{\Te} \}  = - s 			\, , &\quad
	    \{ s , L _{\Te} \} = c  			\, . 	&(7.6) \cr}
$$
So, under the Poisson-brackets, the chosen set of variables forms
the Lie algebra\break
$L ( T ^{\ast} GL (1, \R) \times T ^{\ast} GL (1, \R) \times E _2)$
where $T ^{\ast} GL (1,\R) \equiv
\R \sproduct \plu$, $\sproduct $ is the semi-direct product,
and $E _2 = \R ^2 \sproduct SO(2)$ is the Euclidean group in
two dimensions [14].
Our next step consists of finding a unitary irreducible representation
of the Lie algebra (7.4-6). We choose as our representation space
the space of distributions
$\psi (\Pi _+, \Pi _-, \Te)$ over ${\cal L} ^+ _{(+,+)} \times S ^1$,
and define on it a set of operators $({\what \Pi} _+, {\what \Pi} _-,
{\what c} , {\what s} , {\what L _+} , {\what L _-}, {\what L _{\Te}})$
such that their only non-vanishing
commutators correspond to ${\rm i} {\cet}$ times the Poisson-brackets
(7.4-6).  The action of these operators on
$\psi (\Pi _+, \Pi _-, \Te)$ can be consistently defined in the form
$$
\eqalignno { \Bigl ( {\what \Pi} _{\pm} \psi \Bigr )
(\Pi _+, \Pi _-, \Te) &= \Pi _{\pm} \psi (\Pi _+, \Pi _-, \Te)
						\, \, \, , 	&(7.7) \cr
              \Bigl ( {\what c} \psi \Bigr )
( \Pi _+,  \Pi _-, \Te) &= \cos \Te \psi ( \Pi _+,  \Pi _-, \Te)
						\, \, \, ,	&(7.8) \cr
              \Bigl ( {\what s} \psi \Bigr )
(\Pi _+, \Pi _-, \Te) &= \sin \Te \psi (\Pi _+, \Pi _-, \Te)
						\, \, \, , 	&(7.9) \cr
               	\Bigl ( {\what L _+} \psi \Bigr )
(\Pi _+, \Pi _-, \Te) &= - {\rm i}{\cet } {\Pi _+}
{\partial \over {\partial \Pi _+}} \psi (\Pi _+, \Pi _-, \Te)
						\, \, \, , 	&(7.10) \cr
      	     \Bigl ( {\what L _-} \psi \Bigr )
(\Pi _+, \Pi _-, \Te) &= - {\rm i}{\cet } {\Pi _-}
{\partial \over {\partial \Pi _-}} \psi (\Pi _+, \Pi _-, \Te)
						\, \, \, , 	&(7.11) \cr
             \Bigl ( {\what L _{\Te}} \psi \Bigr )
(\Pi _+, \Pi _-, \Te) &= - {\rm i}{\cet }
{\partial \over {\partial \Te}} \psi (\Pi _+ \Pi _-, \Te)
						\, \, \, . 	&(7.12) \cr}
$$
The inner product in the space of quantum physical states
can then be expressed as:
$$
\eqalignno { < \phi | \psi > &= \int _{\cal Q} \mu \, {\bar \phi} {\psi}
						\, \, \, , 	&(7.13) \cr}
$$
where ${\cal Q} \equiv {\cal L} ^+ _{(+,+)} \times S ^1$, and the measure
$\mu$ is given by
$$
\eqalignno { \mu &= {d\Pi _+\over \Pi _+} {d\Pi _-\over \Pi _-} d \Te
						\, \, \, . 	&(7.14) \cr}
$$
The measure $\mu$ is simply the product of the measures that correspond
to the unitary irreducible representations of the groups
$T ^{\ast} GL (1, \R)$ (twice) and $E _2$ [14].
It is straightforward to check that our operators are self-adjoint
with respect to the scalar product (7.13). Furthermore, there is no
nontrivial subspace in our representation space which remains
invariant under the action of the operators (7.7-12). This proves
that the constructed representation of the algebra (7.4-6)
is unitary and irreducible.

	Let us consider now the type I Bianchi model. In this case, the
reduced configuration space can be identified with
${\cal L} ^+_{(+,+)} \times S ^2 \times S ^1$ (with certain restrictions
still to be imposed in the space of physical states that come from the
identification of points in (6.15)). The sphere $S^2$ is
parametrized by the angles
${\twid \a}$ and $\b ( \b \in [ 0 , \pi ) , {\twid \a} \in [ 0 , 2\pi ) )$,
the circle $S ^1$ by the angle $\te$, and ${\cal L} ^+_{(+,+)}$
is defined by the constraint $p _0 = \sqrt {{p_+} ^2 + {p_-} ^2}$,
with $p_+,p_-,p_0 \in \plu $. From now on, we adopt the compact
notation $p = ( p _+ , p_- )$, $\g=({\twid \a}, \b , \te )$.
Given the form of the reduced configuration space, it is natural to choose
the following over-complete set of configuration variables:
$$
\eqalignno { ( p_+ , p_- , c , & s , k_1 , k_0 , k_{-1} )
						\, \, \, , 	&(7.15) \cr}
$$
where
$$
\eqalignno { c &\equiv \cos \te \, , \quad s \equiv \sin \te
							\, ,   &(7.16a) \cr
k_1 \equiv Y _1^1 ({\twid \a}, \b) \, , &\quad
k_0 \equiv Y _1^0 ({\twid \a}, \b) \, , \quad
k_{-1} \equiv Y _1^{-1} ({\twid \a}, \b) 		\, , 	&(7.16b) \cr}
$$
and $Y _l^m({\twid \a}, \b)$ are the spherical harmonics on
the two-sphere [25]. All the functions in (7.15) are real,
except $k_1$ and $k_{-1}$, which satisfy
$$
\eqalignno { (k _1) ^{\ast} ({\twid \a} , \b ) &=
            - k _{-1} ({\twid \a} , \b ) 		\, .	&(7.17) \cr}
$$
In this case, our generalized momentum variables are
$$
\eqalignno { &L _+ = p_+ U + {{p_+} ^2 \over p_0} T	 \,  ,  \quad
L _- = p_- V + {{p_-} ^2 \over p_0} T 			  \, , 	 &(7.18) \cr
            L _{\twid \a}  = {\twid F _1}  	 	& \, ,  \quad
L ^{\pm} _{({\twid \a}, \b)} = e ^{{\pm}{\rm i}{\twid \a}}
\Bigl ( {\pm} {\rm i} F _2 - \cot \b \,  {\twid F _1} \Bigr )
							   \, ,	 \quad
		   L _\te = F _3			   \, ,  &(7.19) \cr}
$$
with $( T , U , V , {\twid F _1} , F _2 , F _3 )$ the momenta
canonically conjugate to
$(p _0 , p_+ , p _- , {\twid \a} , \b , \te)$.
Following a similar discussion to that presented for Bianchi II,
one can check that our set of variables forms the Lie algebra
$L ( T ^{\ast} GL (1, \R) \times T ^{\ast} GL (1, \R) \times E _3
\times E _2)$ (with respect to the Poisson-brackets structure),
where $E _3 = \R ^3 \sproduct SO(3)$ is the Euclidean group
in three dimensions [14]. We now proceed to find a unitary irreducible
representation of the corresponding Lie algebra of Dirac observables
on the space of distributions $\psi ( p , \g )$
over the reduced configuration space ${\cal L} ^+_{(+,+)} \times
S ^2 \times S ^1$. Parallel to the situation in type II,
the configuration operators that correspond to the variables (7.15)
act as multiplicative operators on the chosen representation
space, while the action of the momentum operators (7.18,19)
can be defined by
$$
\eqalignno { \Bigl ( {\what L _+} \psi \Bigr ) ( p , \g)
&= - {\rm i}{\cet } p _+ {\partial \over {\partial p _+}} \psi ( p , \g )
						\, \, \, , 	&(7.20) \cr
      	     \Bigl ( {\what L _-} \psi \Bigr ) ( p , \g )
&= - {\rm i}{\cet } p _- {\partial \over {\partial p _-}} \psi ( p , \g )
						\, \, \, , 	&(7.21) \cr
             \Bigl ( {\what L _{\te}} \psi \Bigr ) ( p , \g )
&= - {\rm i}{\cet } {\partial \over {\partial \te}} \psi ( p , \g)
						\, \, \, , 	&(7.22) \cr
             \Bigl ( {\what L _{\twid \a}} \psi \Bigr ) ( p , \g )
&= - {\rm i}{\cet } {\partial \over {\partial {\twid \a}}} \psi ( p , \g)
                                                 \, \, \, , 	&(7.23) \cr
\Bigl ( {\what L} ^{\pm} _{({\twid \a}, \b)} \psi \Bigr ) ( p , \g )
&={\cet } e ^{{\pm}{\rm i}{\twid \a}}
\Bigl ( {\pm} {\partial \over {\partial \b}} + {\rm i} \cot \b \,
{\partial \over {\partial {\twid \a}}} \Bigr ) \psi ( p , \g)
                                                 \, \, \, . 	&(7.24) \cr}
$$
The only non-vanishing commutators in this algebra are
$$
\eqalignno {[ {\what {\Pi}}_+ , {\what L _+} ] = {\rm i} {\cet }
{\what {\Pi}}_+					\, , 	&	\quad
            [ {\what {\Pi}}_- , {\what L _-} ] = {\rm i} {\cet }
{\what {\Pi}}_- 				\, , 	&(7.25)\cr
            [ {\what c} , {\what L _{\te}} ] =  - {\rm i} {\cet }
{\what s}				        \, ,	&	\quad
	    [ {\what s} , {\what L _{\te}} ] = {\rm i} {\cet }
{\what c}		                        \, , 	&(7.26) \cr
	    [ {\what L _{\twid \a}} , {\what L ^{\pm}} _{({\twid \a}, \b)}]
= {\pm} {\cet } {\what L ^{\pm}} _{({\twid \a}, \b)}
						\, , 	& 	\quad
	    [ {\what L ^+} _{({\twid \a}, \b)} ,
{\what L ^-} _{({\twid \a}, \b)}] = 2 {\cet } {\what L_{\twid \a}}
						\, , 	&(7.27) \cr
	    [ {\what L _{\twid \a}} , {\what k _{\pm 1}}] = {\pm}
{\cet } {\what k _{\pm 1}}	  	\, , 	& 	\quad
	    [ {\what L ^{\pm}} _{({\twid \a}, \b)} , {\what k _m} ]
= {\cet } \sqrt {2 - m (m \pm 1) } \, {\what k _m}
					\, , 		&(7.28) \cr}
$$
where $m = 1 , 0 , -1$. The inner product in the space of
quantum states, without imposing the quantum analogue to the
identification of points given by (6.15), takes on
the expression
$$
\eqalignno { < \phi | \psi > &= \int _{\cal Q} \mu {\bar \phi} \psi
						\, , 	&(7.29) \cr}
$$
where ${\cal Q} = {\cal L} ^+_{(+,+)} \times S ^2 \times S ^1$,
and the measure $\mu$ is
$$
\eqalignno { {\mu } &= {dp_+\over p_+} {dp_-\over p_-}
\sin \b  d\b d{\twid \a} d\te		\, \, \, . 	&(7.30) \cr}
$$
Note that all the Dirac observables are represented
by self-adjoint operators, except ${\what k _m}$ and
${\what L ^{\pm}}_{({\twid \a}, \b)}$, which satisfy
$$
\eqalignno { &({\what k _m} )^{\dagger} = (-1) ^m {\what k _{-m}}
					\, \, \, , 	&(7.31) \cr
&({\what L ^+} _{({\twid \a}, \b)}) ^{\dagger} =
{\what  L ^-} _{({\twid \a}, \b)}       \, \, \, . 	&(7.32) \cr}
$$
Let us impose now the identification of points (6.15) in the reduced phase
space as a restriction on the physical states $\psi \in L ^2 \Bigl (
{\cal L} ^+_{(+,+)} \times S ^2 \times S ^1 , \mu \Bigr )$. In the
representation that we have chosen, the required restriction can be stated as
$$
\eqalignno { \psi ( p , {\twid \a} , \b , \te ) &=
\psi ( p , - {\twid \a}, \pi - \b , \te + \pi )
					\, \, \, . 	&(7.33) \cr}
$$
The operators that correspond to the canonical variables
${\twid F_1}$ and $F_2$ in the reduced phase space are clearly given,
in our representation, by
${\what L _{\twid \a}}$ and ${\what L _{\te}}$, respectively.
It is easy to check, using Eq. (7.33), that
$$
\eqalignno { \Bigl ( {\what L _{\twid \a}} \psi \Bigr )
( p , {\twid \a} , \b , \te ) &= - \Bigl ( {\what L _{\twid \a}}
\psi \Bigr ) ( p , - {\twid \a} , \pi - \b , \te + \pi )
	     				\, \, \, , 	&(7.34) \cr}
$$
and similarly
$$
\eqalignno { \Bigl ( {\what L _{\te}} \psi \Bigr )
( p , {\twid \a} , \b , \te ) &= - \Bigl ( {\what L _{\te}}
\psi \Bigr ) ( p , - {\twid \a} , \pi - \b , \te + \pi )
	                                 \, \, \, .	&(7.35) \cr}
$$
On the other hand, and apart from the factor ordering ambiguities that must be
irrelevant in the classical limit, the variable $F_2$ in the reduced
phase space  can be represented by the symmetrized operator corresponding to
$L _{\b} = - {\rm i} (e ^{-{\rm i} {\twid \a}} L _+ + \cot \b \,
L _{\twid \a} )$, where
$e ^{{\rm i} {\twid \a}}$  and $\cot \b$ must be expressed in terms of
$k _m$, $m=1,0,-1$. It is then possible to show that
$$
\eqalignno { \Bigl ( {\what L _{\b}} \psi \Bigr )
( p , {\twid \a} , \b , \te ) &= - \Bigl ( {\what L _{\b}} \psi \Bigr )
( p , - {\twid \a} , \pi - \b , \te + \pi ) + o ({\cet})
					\, \, \, . 	&(7.36) \cr}
$$
Eqs. (7.33-36) guarantee that, in the classical limit, the points
(6.15) are identified in the reduced phase space, recalling that
${\twid \a} = 2 \a$ and, thus, ${\twid F _1} = {\half} F _1$.

	Therefore, the space of quantum physical states for the type I
Bianchi model is simply the Hilbert subspace of functions in
$L ^2 \Bigl ( {\cal L} ^+_{(+,+)} \times S ^2 \times S ^1 ,
\mu \Bigr )$ that satisfy relation (7.33). Using Eqs. (7.29,30)
and (7.33), the inner product in this space is easily computed to be
$$
\eqalignno { < \phi | \psi > &= 2 \int _{\plu} {dp _+\over p _+}
\int _{\plu} {dp _-\over p _-} \int ^{\pi}_0 \! \sin \b d \b
\int ^{2\pi}_0 \! d {\twid \a} \int ^{\pi}_0 \! d \te
{\bar \phi} (p ,\g) \psi (p,\g)			\, , 	&(7.37) \cr}
$$
so that we can restrict our attention to the sector $\te \in [0,\pi)$, and
identify the space of physical states with $L ^2 \Bigl ( {\cal L} ^+_{(+,+)}
\times S ^2 \times [ 0 , \pi) , \mu \Bigr )$. The restriction to this
Hilbert space of the set of operators previously defined in
$L ^2 \Bigl ( {\cal L} ^+_{(+,+)} \times S ^2 \times S ^1 , \mu \Bigr )$
have then a well-defined action in the quantum physical
states for Bianchi type I.

	We have thus succeeded in quantizing the full type I and II
Bianchi models following the non-perturbative canonical approach. The
physical interpretation of the quantum theories so-constructed is completely
analogue to that presented in Ref. [9] for the diagonal Bianchi models.
The results will be discussed elsewhere [26].

\noindent
{\bf VIII. Bianchi Type I: An Alternative Quantization}
\medskip
\noindent
We want to discuss now a different approach to the quantization of the
type I Bianchi model, making use of the symmetries that are present in the
scalar constraint at the classical level [9]. We will show that the quantum
theory obtained in this way is equivalent to the reduced phase space
quantization of this Bianchi model.

	We will restrict our attention to the sector of positive definite
metrics. These metrics can be represented by real triads in $GL(3,\R)$,
which can be uniquely written in the form
$$
\eqalignno { {E ^I} _a &= {(M _T) ^I} _J {(M _D) ^J} _K R ^{aK} (\g _L)
						\, \, \, , 	&(8.1) \cr}
$$
with ${R ^{aK}} (\g _L) \in SO(3) , \, \g _L \, (L = 1, 2, 3)$ the associated
Euler angles (see Eq.(6.11)), and $M_T$ and $M_D$, respectively,
an upper-triangular and a diagonal matrix.
Let us introduce the following basis of generators for
the upper-triangular and diagonal groups:
$$
\eqalignno { {(T ^1) _I} ^J = \d _I^1 \d _2^J	 \, , \quad
	    {(T ^2) _I} ^J &= \d _I^1 \d _3^J	 \, , \quad
             {(T ^3) _I} ^J = \d _I^2 \d _3^J	 \, , 	&(8.2) \cr
{(T ^4) _I} ^J = \d _I^J - 3 \d ^3_I \d ^J_3	 \, , \quad
{(T ^5) _I} ^J &= {\sqrt 3} (\d _I^1 \d _1^J - \d ^2_I \d ^J_2)
						 \, , \quad
{(T ^6) _I} ^J = 2 \d _I^J 	 		 \, . 	&(8.3) \cr}
$$
Then, the matrices $M_T$ and $M_D$ can be uniquely expressed in terms of the
exponentials of these generators as
$$
\eqalignno { M _T = \Pi _{i=1}^3 e ^{ x_i T ^i} \, , &\quad
	     M _D = \Pi _{i=4}^6 e ^{ x_i T ^i} \, . 	&(8.4) \cr}
$$
We can consider $x_i , i = 1 , ... , 6$, and $\g _L , L = 1, 2, 3$,
as a new set of configuration variables. Let us designate their
canonically conjugate momenta by $p ^i$ and ${p ^{\g}} _L$,
and perform a canonical transformation from the Ashtekar variables
$(A_I^a , E^I_a)$ to the new set $(x_i , \g _L , p ^i , {p ^{\g}} _L)$.
Instead of dealing with ${p ^{\g}} _L$, it is more convenient to use the
angular momenta $J ^{\g} _L$ of the Euler angles:
$$
\eqalignno { J ^{\g} _L &= {( Q ^{-1} ) _L} ^K {p ^{\g}} _K
						\, \, \, .	&(8.5) \cr}
$$
The matrix ${Q _L} ^K$ that appears in this equation is defined by means
of the relation [27]
$$
\eqalignno { {\partial {R ^a}_M\over {\partial \g _L}} &=
{\e ^P} _{MN} {R ^a} _P {Q _L} ^N		\, \, \, , 	&(8.6) \cr}
$$
which implicitly employs the fact that $d {R ^a} _M {R ^a} _P$ is an
antisymmetric one-form.

	The spin connection $A_I^a$ can be obtained by integrating the
system of differential equations
$$
\eqalignno { \{ {A _I}^a , {E ^J}_b \} &=
- {\partial {A _I} ^a\over {\partial p^i}}
{\partial {E ^J} _b\over \partial x _i} - {\partial {A _I} ^a\over
{\partial {J ^{\g} _K}}} (Q ^{-1}) _{KL} {\partial {E ^J} _b\over
{\partial {\g _L}}} = {\rm i} \d ^J_I \d ^a_b
						\, \, \, . 	&(8.7) \cr}
$$
The solution to Eq. (8.7) turns out to be a complicated algebraic expression,
although the calculations leading to it are relatively simple. We will proceed
to discuss the conclusions that can be inferred from the result of these
computations without displaying the explicit form of ${A _I} ^a ( x _i,
\g _L, p ^i, {J ^{\g}} _L )$.

	We first note that, since the $SO(3)$ connection ${\G _I} ^a$
vanishes in Bianchi type I, ${A _I} ^a$ is purely imaginary.
As a consequence, all the momenta $(p ^i , {J ^{\g}} _L)$ can be
restricted to be real. On the other hand, substituting ${E ^I} _a$,
given by (8.1-4), and \break
${A _I} ^a ( x _i, \g _L, p ^i, {J ^{\g}} _L )$
in Eq. (2.9), the Gauss law constraints for Bianchi type I in
the introduced set of canonical variables can be rewritten
$$
\eqalignno { {\cal G} _a &= {{\rm i}\over 2} \e _{abc}
R ^{bI} R ^{cJ} {J ^{\g}} _L  \, {\e _{IJ}} ^L \approx 0
						\, \, \, , 	&(8.8) \cr}
$$
{}from which we conclude that
$$
\eqalignno {  {J ^{\g}} _L &\approx 0 		\, \, \, . 	&(8.9) \cr}
$$
The vector constraints (2.10) are empty for the type I Bianchi model. We are
thus left only with the scalar constraint corresponding to (2.11):
$$
\eqalignno { {\cal S} &= \Bigl ( \Tr AE \Bigr ) ^2 -
\Tr \Bigl ( AE \Bigr ) ^2			\, \, \, .	&(8.10) \cr}
$$
Substituting $( {A _I} ^a , {E ^I} _a )$ as functions of $x _i , \g _L$
and $p ^i$ in (8.10) (with ${J ^{\g}} _L$ set equal to zero), we arrive
at the expression
$$
\eqalignno { {\cal S} &= - {1\over 6} (p ^6) ^2 + {1\over 6} (p ^4) ^2 +
{1\over 6} (p ^5) ^2							\cr
+ 2 \Bigl ( e ^{6x _4 - 2{\sqrt 3} x _5} (p ^3) ^2 &+
e ^{6x _4 + 2{\sqrt 3} x _5} (p ^2) ^2 + e ^{4{\sqrt 3} x _5}
(p ^1 - x_3 p ^2) ^2 \Bigr ) 			\, \, \, . 	&(8.11) \cr}
$$
Eq. (8.11) can be considered as a quadratic constraint on the cotangent bundle
over the six dimensional space coordinatized by the variables $x_i$.
We point out that, from Eq. (8.1), this space can be identified with
the space of positive definite metrics, since
$$
\eqalignno { g ^{IJ} (x _i) &= ( \de E ) ^{-1} {E ^I} _a {E ^J} ^a 	\cr
= e ^{-6x_6} \, {(M_T) ^I} _K & {(M_D) ^K} _L {(M_D) ^P} _L {(M_T) ^J} _P
						\, \, \, . 	&(8.12) \cr}
$$
The quadratic form that appears in (8.11) endows this space with the natural
metric [22]:
$$
\eqalignno { d s ^2 &= - 6 (dx _6)^2 + G ^{ij} d {\twid x} _i d {\twid x} _j
						\, \, \, , 	&(8.13) \cr}
$$
where ${\twid x} _i = x _i , i = 1, ... , 5$, and
$$
\eqalignno { G ^{ij} = 6 (dx _4) ^2 &+ 6 (dx _5) ^2
+ {\half} \Bigl ( e ^{-6x _4 + 2{\sqrt 3} x _5} (dx _3) ^2 +            \cr
e ^{-4{\sqrt 3} x _5} (dx _1) ^2 &+ e ^{-6x _4 - 2{\sqrt 3} x _5}
(dx _2 + x_3 d x_1) ^2 \Bigr ) 			\, \, \, . 	&(8.14) \cr}
$$
The coordinate $x^6$ plays then the role of a time, and $\partial _6$
is a global time-like Killing vector of the metric (8.13). From Eqs. (8.12) and
(8.2-4), the five dimensional space coordinatized by ${\twid x} _i$ is simply
the space of positive definite metrics of unit determinant. Following now a
completely similar analysis to that  carried out by Henneaux, Pilati and
Teitelboim in Ref. [22], it is possible to prove that this five dimensional
space, provided with the metric (8.14), can be identified with the coset space
${SL (3,\R)\over SO(3)}$ [28]. Furthermore, they showed that, in the metric
representation, the quantum states for Bianchi type I can be
decomposed in the form
$$
\eqalignno { f (g) &= \int _{\R} d {\l} _+ \int _{\R} d {\l} _-
\mu ( \l ) \int _0^{\pi} \! d\a \int _0^{\pi} \! \sin \b d\b
\int _0^{\pi} \! d \te {\twid f} (\l , \g )
e _{\l , R ( \g )} ( g ) 				\, , 	&(8.15) \cr}
$$
where $\l = (\l _+ , \l _- ) , \, \mu ( \l )$ is a specific measure
over $\R ^2$, $( \a , \b , \te ) \equiv (\g _L)$
are the Euler angles that parametrize the matrices
$R (\g _L) \in SO(3)$, and $e _{\l , R (\g )} ( g )$
are generalized ``plane waves'' which satisfy the
operator constraint associated to the classical constraint (8.11).
For any matrix $R (\g _L) \in SO(3), \, e _{\l , R (\g _L)} ( g )$
is defined as the ``plane wave'' of the rotated metric $g$: $\,
e _{\l R (\g)} ( g ) = e _{\l , I} ( R g R ^t )$,
with $e _{\l , I} ( g )$ given in the parametrization (8.2-4) by
the expression
$$
\eqalignno { e _{\l , I} ( g ) &= e ^{-{\rm i} \l _0 (\l) x _6 +
( {\rm i} \l _+ + 3 ) x _4 + ( {\rm i} \l _- + \sqrt 3 ) x _5 }
						\, \, \, .	&(8.16) \cr}
$$
The ``plane waves'' (8.16) are eigenfunctions of the momenta operators
${\what p} _4 , {\what p} _5$ and ${\what p} _6$, defined as self-adjoint
with respect to the metric (8.13,14) [22]. The respective eigenvalues
are ${\l} _+ , {\l} _-$ and ${\l} _0$. In order to fulfill the operator
constraint corresponding to (8.11),
$({\l} _0 , {\l} _+ , {\l} _-  )$ must satisfy the relation
$$
\eqalignno { - {\l _0} ^2 + {\l _+} ^2 + {\l _-} ^2 &= 0
						\, \, \, , 	&(8.17) \cr}
$$
which is a direct analogue of the scalar constraint (3.6).
In fact, it is possible
to check (from Eqs. (8.2-4), (8.12), and (3.4)) that the set of
parameters $( \b _0 , \b _+ , \b _-)$ employed in Sec. III to describe the
diagonal metrics for the type I model are related to the coordinates
$(x ^4 , x ^5 , x^6)$ by means of the transformation
$$
\eqalignno {\!
\b _0 &= {1\over {\sqrt 3}} (2 x^6 + x ^4) + C _1 \, , \,
\b _+  = {1\over {\sqrt 3}} (x^6 + 2 x ^4) + C _2  \, , \,
\b _-  = - x ^5 + C _3			\, , 	&(8.18) \cr}
$$
with $C _1 , C _2 , C _3$ some unspecified constants.
The canonically conjugated variables to the $\b $'s, $(p _0, p _+, p _-)$,
used throughout our analysis of the Bianchi type I, can then be obtained from
$({\l} _0 , {\l} _+ , {\l} _- )$ by completing the
canonical transformation (8.18):
$$
\eqalignno { p _0 &= {1\over {\sqrt 3}} (2 \l _0 +   \l _+)  \, , \quad
             p _+  = {1\over {\sqrt 3}} (- \l _0 + 2 \l _+)  \, , \quad
             p _-  = - \l _-  				\, . 	&(8.19) \cr}
$$
If we want to consider exclusively inequivalent spacetime geometries,
and not only different positive definite spatial metrics, we have to
restrict the range of $p _0, p_+$ and $p_-$ to lie in the positive
real axis, as we proved in Sec. VI.
Changing coordinates from $({\l} _0 , {\l} _+ , {\l} _- )$ to
$(p _0, p _+, p _-)$, imposing that $p_0 = \sqrt { {p _+} ^2 + {p _-} ^2}$,
and restricting \break
$p _+, p _- \in \plu$, we arrive, from (8.15),
at the following decomposition for the quantum sates of the
type I spacetime geometries:
$$
\eqalignno { F (g) &= \int _{\Lig} \! d p _+ d p _-  {\twid \mu} ( p )
\int _0^{\pi} \! d\a \int _0^{\pi} \! \sin \b d\b
\int _0^{\pi} \! d \te {\twid F} ( p , \g )  e _{p, R (\g )} ( g )
 							\, , 	&(8.20) \cr}
$$
where $p \equiv ( p _+ , p _- ) , {\twid \mu} (p)$ is certain measure
over $\Lig$ and $e _{p , R (\g )} ( g )$ are the generalized
``plane waves'' expressed in terms of $p _+ , p _-$ and $\g $.
The wave function ${\twid F} ( p , \g )$
characterizes uniquely the quantum state
$F (g)$ [22], so that, instead of the metric
representation, we can select the representation
$( p , \g )$ for the quantization of the model.
The decomposition (8.20) induces the following
inner product:
$$
\eqalignno {\! \! \! < {\twid F} , {\twid G} > &= \int _{\Lig} \!
d p _+ d p _- {\twid \mu} ( p ) \int _0^{\pi} \! d \a \int _0^{\pi}
\sin \b d\b \int _0^{\pi} \! d \te {\twid F} ^{\ast} ( p , \g )
{\twid G} ( p , \g ) 				\, . 	&(8.21) \cr}
$$
It is obvious that the representation $( p , \g )$ coincides with that
used in Sec. VII for the quantization of Bianchi I. Furthermore,
the inner product (8.21) in the space of quantum states differs
only in the choice of the measure ${\twid \mu} ( p )$ from that
determined in (7.37). As a consequence, we conclude that the two quantum
theories here discussed for Bianchi type I result
to be unitarily equivalent.

\noindent
{\bf IX. Conclusions}
\medskip
\noindent
We have succeeded in completing the canonical quantization
of the type I and II Bianchi models while keeping the totality of
degrees of freedom of these homogeneous gravitational minisuperspaces.
Our analysis generalizes the works existing so far in the literature [9,10],
which had concentrated their attention on the diagonal reduction of these
systems.

	We have first calculated the explicit expressions of the
general solution for these two Bianchi types, using the fact
that the classical evolution problem can always be brought to
the diagonal form by a change in the set of left-invariant one-forms
on the leaves of the homogeneous foliation.
This is the first time, to our knowledge, that the general solution
for Bianchi II has been explicitly displayed.

	The classical solutions have been written in both
geometrodynamic and Ashtekar variables, restricted to the sector of
positive definite metrics. We have determined the
sets of solutions that correspond to different spacetime
geometries, eliminating the overcounting of physical states.
For both types I and II, the parameters used to describe the relevant
non-degenerate physical solutions have been shown to define
a good (analytic) set of coordinates in the phase spaces of these
models. The presence of the non-diagonal degrees of freedom
plays an essential role in the proof of this statement. If we
had performed a similar analysis in the reduced diagonal models,
the phase space coordinates associated to the diagonal degrees
of freedom could have been extended to a wider range of analyticity,
because, in this case, one can consistently consider as fixed the
set of left-invariant one-forms in the homogeneous foliation. The
physical states of the quantum theories constructed thereafter for the
diagonal models can be interpreted as dependent on three-geometries
with some preferred directions. In our description, however,
one is forced to treat the directions in the homogeneous
slices as being interchangeable, as far as this is allowed by
the symmetries of the Bianchi model. Apart form the
maintenance of the extra degrees of freedom, this is the main difference
that arises in the study of the full Bianchi types I and II with
respect to the analysis of their diagonal counterparts.
The constraints of the systems are identical for both kinds of models,
since the whole classical evolution can always be
put into diagonal form.

	Performing a transformation from
the Ashtekar variables to the sets of phase space coordinates introduced
for Bianchi I and II, we have endowed the reduced phase space of each of
these minisuperspaces with an analytic symplectic structure.
For the type II Bianchi model, the symplectic form obtained in this
way can be identified with that of the real cotangent bundle over
the reduced configuration space ${\Lig} \times S ^1$; for type I,\break
the symplectic form corresponds to the real cotangent bundle over
${\Lig} \times S ^2 \times S ^1$, with an additional identification of points
in $S ^2 \times S ^1$ that we have chosen to impose at the quantum level.

	We have proceeded to quantize the models by selecting
a complete and closed \break
$\ast -$ algebra of Dirac observables in
the reduced phase space of each of these systems. We have then constructed
an explicit unitary irreducible representation of that algebra
on the space of distributions in the reduced configuration
space. In this way, the \break
${\ast} -$ relations in the algebra
of observables have been
straightforwardly implemented in the adopted representation as adjointness
relations among the quantum operators and the inner product in
the space of quantum states has been subsequently fixed.
The physical interpretation of the mathematical framework
obtained here is essentially the same as that presented in Ref. [9] for
the diagonal reduction of the Bianchi models.

	Finally, we have outlined a different approach to
the quantization of Bianchi type I, \break
using the existence of a
conditional symmetry [29] to span the quantum states in terms of
generalized ``plane waves''. This alternative approach may be useful
for mini\-superspace models whose algebra of Dirac observables is not
sufficiently known, so that the canonical quantization program cannot be
yet applied to completion in these cases. For Bianchi I, the two quantum
theories analyzed in this work have been shown to be unitarily
equivalent.

	We have thus provided the full type I and II Bianchi models
with the framework needed to address cosmological problems from
the quantum point of view. Our study also illuminates the role played by
the additional non-diagonal degrees of freedom, which turn out to
be frozen in the Hamiltonian description. Let us finally point out
that, from our discussion in Sec. VII, the Hilbert subspace of states
that are separable as a function in $\Lig$ times a function in
the ``non-diagonal'' part of the reduced configuration
space can be immediately associated with quantum physical states for
the corresponding diagonal case, by simply neglecting the
dependence on the non-diagonal configuration variables.
One can thus analyze the implications of the
quantum reduction of degrees of freedom for these two kinds of
minisuperspace models, as toy models in which one can check the possible
validity of the minisuperspace approximation in the full quantum
theory of gravity [30].

\noindent
{\bf Acknowledgements}
\medskip
\noindent
The authors are greatly thankful to A. Ashtekar and J. Louko
for helpful discussions and useful comments, and to D. Marolf
for suggesting some corrections to the original manuscript.
N. Manojlovi\' c would also like to thank C. Isham, who
initially proposed the subject of this work.
He gratefully acknowledges the financial support provided by the
Relativity Group at Syracuse University, as well as their
warm hospitality.\break
G. A. Mena Marug\' an was supported by the Spanish Ministry
of Education and Science Grant No. EX92 06996911. He is grateful
for the hospitality and partial support of the Department
of Physics at Syracuse University.

\bigskip

\noindent
{\bf References}
\medskip
\frenchspacing
\item {[1]} A. Ashtekar, Phys. Rev. Lett. {\bf 57}, 2244 (1986);
Phys. Rev. D {\bf 36}, 1587 (1987).
\item {[2]} A. Ashtekar, J. D. Romano and R. S. Tate, Phys. Rev. D {\bf 40},
2572 (1989).
\item {[3]} A. Ashtekar, {\sl Lectures on Non-Perturbative Canonical Gravity}
(World Scientific, Singapore, 1991), and references therein.
\item {[4]} A. Ashtekar, in {\sl The Proceedings of the 1992 Les Houches
School on Gravitation and Quantization}, edited by B. Julia
(North-Holland, Amsterdam, to be published).
\item {[5]} A. Rendall, Syracuse University Report No. SU-GP-93/2-2,
1993 (unpublished).
\item {[6]} I. Bengtsson, Class. Quantum Grav. {\bf 5}, L139 (1988);
{\sl ibid} {\bf 7}, 27 (1989);
S. Koshti and N. Dadhich, Class. Quantum Grav. {\bf 6}, L223 (1989);
L. Bombelli and R. J. Torrence, Class. Quantum Grav. {\bf 7}, 1747 (1990);
H. Kastrup and T. Thiemann, Nucl. Phys. {\bf B} (to be published).
\item {[7]} H. Kodama, Prog. Theor. Phys. {\bf 80}, 1024 (1988);
Phys. Rev. D {\bf 42}, 2548 (1990).
\item {[8]} V. Moncrief and M. Ryan, Phys. Rev. D {\bf 44}, 2375 (1991).
\item {[9]} A. Ashtekar, R. Tate, and C. Uggla, {\sl Minisuperspaces:
Observables and Quantization}, Syracuse University Report, 1993
(unpublished).
\item {[10]} G. Gonz\' alez and R. Tate, Syracuse University Report, 1993
(unpublished).
\item {[11]} A. Ashtekar and J. Pullin, Ann. Israel Phys. Soc. {\bf 9},
65 (1990).
\item {[12]} M. P. Ryan and L. C. Shepley, {\sl Homogeneous Relativistic
Cosmologies} (Princeton University Press, Princeton, 1975).
\item {[13]}  R. M. Wald, {\sl General Relativity} (University of Chicago
Press, Chicago, 1984).
\item {[14]} C. Isham, {\sl Topological and Global Aspects of Quantum
Theory}, Les Houches Session XL, 1983, edited by B. S. DeWitt and
R. Stora (Elsevier Science Publishers B. V., 1984).
\item {[15]} A. Ashtekar and J. Samuel, Class. Quantum Grav. {\bf 8},
2191 (1991).
\item {[16]} D. Kramer, H. Stephani, M. MacCallum, and E. Herlt,
{\sl Exact Solutions of Einstein's Field Equations} (Cambridge University
Press, Cambridge, England, 1980).
\item {[17]} N. Manojlovi\'c and A. Mikovi\'c, Class. Quantum Grav. {\bf 10},
559 (1993).
\item {[18]} R. T. Jantzen, in {\sl Cosmology of the Early Universe},
edited by L. Z. Fang and R. Ruffini (World Scientific, Singapore, 1984);
K. Rosquist, C. Uggla, and R. T. Jantzen, Class. Quantum Grav. {\bf 7},
611 (1990).
\item {[19]} R. T. Jantzen, Commun. Math. Phys. {\bf 64}, 211 (1979).
\item {[20]} A. Helfer, M. S. Hickman, C. Kozameh, C. Lucey and
E. T. Newman, Gen. Rel. Grav. {\bf 20}, 875 (1988).
\item {[21]} C. W. Misner, in {\sl Magic Without Magic}, edited by
J. Klauder (Freeman, San Francisco, 1972).
\item {[22]} M. Henneaux, M. Pilati and C. Teitelboim, Phys. Lett.
{\bf 110B}, 123 (1982).
\item {[23]} L. V. Ahlfors, {\sl Complex Analysis} (McGraw-Hill, New York,
1979).
\item {[24]} G. A. Mena Marug\' an, Phys. Rev. D {\bf 42}, 2607 (1990);
{\sl ibid} {\bf 46}, 4320 (1992).
\item {[25]} See, for instance, A. Messiah, {\sl Quantum Mechanics},
Vol. I (North-Holland, Amsterdam, 1961).
\item {[26]} N. Manojlovi\' c and G. A. Mena Marug\' an (in preparation).
\item {[27]} J. Goldstone and R. Jackiw, Phys. Lett. {\bf 74B}, 81 (1978).
\item {[28]} B. S. DeWitt, Phys. Rev. {\bf 160}, 1113 (1967).
\item {[29]} K. Kucha\v r, J. Math. Phys. {\bf 22}, 2640 (1982).
\item {[30]} K. Kucha\v r and M. P. Ryan, Jr., in {\sl Gravitational
Collapse and Relativity}, edited by H. Sato and T. Nakamura (World
Scientific, Singapore, 1986); Phys. Rev. D {\bf 40}, 3982 (1989).

\bye